\documentclass[10pt,a4paper]{article}
\usepackage{geometry}
\usepackage{amsmath,amssymb}
\usepackage{changepage}
\usepackage[utf8x]{inputenc}
\usepackage{textcomp,marvosym}
\usepackage{cite}
\usepackage{hyperref}
\usepackage[right]{lineno}
\usepackage{microtype}
\DisableLigatures[f]{encoding = *, family = * }
\usepackage[table]{xcolor}
\usepackage{array}
\newcolumntype{+}{!{\vrule width 2pt}}
\newlength\savedwidth

\usepackage{setspace} 
\raggedright
\usepackage[aboveskip=1pt,labelfont=bf,labelsep=period,justification=raggedright,singlelinecheck=off]{caption}
\renewcommand{\figurename}{Fig}
\makeatletter
\renewcommand{\@biblabel}[1]{\quad#1.}
\makeatother
\date{}
\usepackage{lastpage,fancyhdr,graphicx}
\usepackage{epstopdf}
\usepackage{bm}
\renewcommand{\vec}[1]{\boldsymbol{\mathbf{#1}}}
\newcommand{\dd}{{\rm d}}
\renewcommand{\eqref}[1]{Eq~(\ref{#1})}
\newcommand{\figref}[1]{\figurename~\ref{#1}}

\usepackage{etoolbox}
\newtoggle{showFigs}
\toggletrue{showFigs}
\newtoggle{showSuppFigs}
\toggletrue{showSuppFigs}

\begin{document}
\vspace*{0.2in}

\begin{flushleft}
{\Large
\textbf\newline{Clonal Dominance and Transplantation Dynamics in Hematopoietic Stem Cell Compartments} 
}
\newline
\\
Peter Ashcroft\textsuperscript{1*},
Markus G. Manz\textsuperscript{2},
Sebastian Bonhoeffer\textsuperscript{1}
\\
\bigskip
\textbf{1} Institut f\"ur Integrative Biologie, ETH Z\"urich, Universit\"atstrasse 16, CH-8092 Z\"urich, Switzerland
\\
\textbf{2} Division of Hematology, University Hospital Z\"urich and University of Z\"urich, Schmelzbergstrasse 12, CH-8091 Z\"urich, Switzerland
\\
\bigskip

* peter.ashcroft@env.ethz.ch
\end{flushleft}


\section*{Abstract}
Hematopoietic stem cells in mammals are known to reside mostly in the bone marrow, but also transitively passage in small numbers in the blood.
Experimental findings have suggested that they exist in a dynamic equilibrium, continuously migrating between these two compartments.
Here we construct an individual-based mathematical model of this process, which is parametrised using existing empirical findings from mice.
This approach allows us to quantify the amount of migration between the bone marrow niches and the peripheral blood.
We use this model to investigate clonal hematopoiesis, which is a significant risk factor for hematologic cancers.
We also analyse the engraftment of donor stem cells into non-conditioned and conditioned hosts, quantifying the impact of different treatment scenarios.
The simplicity of the model permits a thorough mathematical analysis, providing deeper insights into the dynamics of both the model and of the real-world system.
We predict the time taken for mutant clones to expand within a host, as well as chimerism levels that can be expected following transplantation therapy, and the probability that a preconditioned host is reconstituted by donor cells.

\section*{Author Summary}
Clonal hematopoiesis -- where mature myeloid cells in the blood deriving from a single stem cell are over-represented -- is a major risk factor for overt hematologic malignancies.
To quantify how likely this phenomena is, we combine existing observations with a novel stochastic model and extensive mathematical analysis.
This approach allows us to observe the hidden dynamics of the hematopoietic system.
We conclude that for a clone to be detectable within the lifetime of a mouse, it requires a selective advantage.
I.e. the clonal expansion cannot be explained by neutral drift alone.
Furthermore, we use our model to describe the dynamics of hematopoiesis after stem cell transplantation.
In agreement with earlier findings, we observe that niche-space saturation decreases engraftment efficiency.
We further discuss the implications of our findings for human hematopoiesis where the quantity and role of stem cells is frequently debated.

\section*{Introduction}
The hematopoietic system has evolved to satisfy the immune, respiratory, and coagulation demands of the host.
A complex division tree provides both amplification of cell numbers and a variety of differentiated cells with distinct roles in the body \cite{kondo:ARI:2003,paul:Cell:2015,kaushansky:book:2016}.
In a typical adult human $\sim 10^{11}$ terminally differentiated blood cells are produced each day \cite{vaziri:PNAS:1994,kaushansky:book:2016,nombela:BloodAdv:2017}.
It has been argued that the division tree prevents the accumulation of mutations, which are inevitable given the huge number of cell divisions \cite{werner:Interface:2013,brenes:Interface:2013,derenyi:NatComms:2017}.
At the base of the tree are hematopoietic stem cells (HSCs).
These have the ability to differentiate into all hematopoietic cell lineages, as well as the capacity to self-renew \cite{passegue:JEM:2005,kondo:ARI:2003}, although the exact role of HSCs in blood production is still debated \cite{sawai:Immunity:2016,schoedel:Blood:2016}.

With an aging population, hematopoietic malignancies are increasingly prevalent \cite{sant:Blood:2010}.
Clonal hematopoiesis -- where a lineage derived from a single HSC is overrepresented -- has been identified as a significant risk factor for hematologic cancers \cite{genovese:NEJM:2014,jaiswal:NEJM:2014,xie:NatMed:2014}.
To assess the risks posed to the host we need an understanding of how fast clones are growing, when they initiate, and if they would subvert physiologic homeostatic control.

The number of HSCs within a mouse is estimated at ${\sim \! 0.01\%}$ of bone marrow cellularity \cite{bhattacharya:JEM:2006,bryder:AJPath:2006}, which amounts to ${\sim \! 10,000}$ HSCs per host \cite{abkowitz:Blood:2002,bhattacharya:JEM:2006,bhattacharya:JEM:2009,kaushansky:book:2016}.
In humans this number is subject to debate; limited data has lead to the hypothesis that HSC numbers are conserved across all mammals \cite{abkowitz:Blood:2002}, but the fraction of `active' HSCs depends on the mass of the organism \cite{dingli:PLoSONE:2006} (see also Refs~\cite{dingli:bookchapter:2009,nombela:BloodAdv:2017} for a discussion).

Within an organism, the HSCs predominantly reside in so-called bone marrow niches: specialised micro-environments that provide optimal conditions for maintenance and regulation of the HSCs \cite{morrison:Nature:2014,crane:NatRevImmun:2017}.
There are likely a finite number of niches within the bone marrow, and it is believed that they are not all occupied at the same time \cite{bhattacharya:JEM:2006}.
The number of niches is likely roughly equal to the number of HSCs, and through transplantation experiments in mice it has been shown that ${\sim \! 1\%}$ of the niches are unoccupied at any time \cite{bhattacharya:JEM:2006,czechowicz:Science:2007}.
A similar number of HSCs are found in the peripheral blood of the host \cite{bhattacharya:JEM:2006}.
These free HSCs are phenotypically and functionally comparable to (although distinguishable from) bone marrow HSCs \cite{wright:Science:2001,bhattacharya:JEM:2009}.
The HSCs have a residence time of minutes in the peripheral blood, and parabiosis experiments (anatomical joining of two individuals) have shown that circulating HSCs can engraft to the bone marrow \cite{wright:Science:2001}.
It has also been shown that HSCs can detach from the niches without cell division taking place \cite{bhattacharya:JEM:2009}.
These findings paint a picture of HSCs migrating between the peripheral blood and the bone marrow niches, maintaining a dynamic equilibrium between the two compartments.

In this manuscript we construct a model from the above described processes, and we use this to answer questions about clonally dominant hematopoiesis.
We first consider this in mice, where we use previously reported values to parametrise our model.
The model is general enough that it also captures scenarios of transplantation into both preconditioned (host HSCs removed) and non-preconditioned hosts: the free niches and the migration between compartments also allows for intravenously injected donor HSCs to attach to the bone marrow niches and to contribute to hematopoiesis in the host.
In the discussion we comment on the implications of these results for human hematopoiesis.

\section*{Materials and Methods}

\begin{figure}[h]
\centering

\iftoggle{showFigs}{\includegraphics[width=0.6\linewidth]{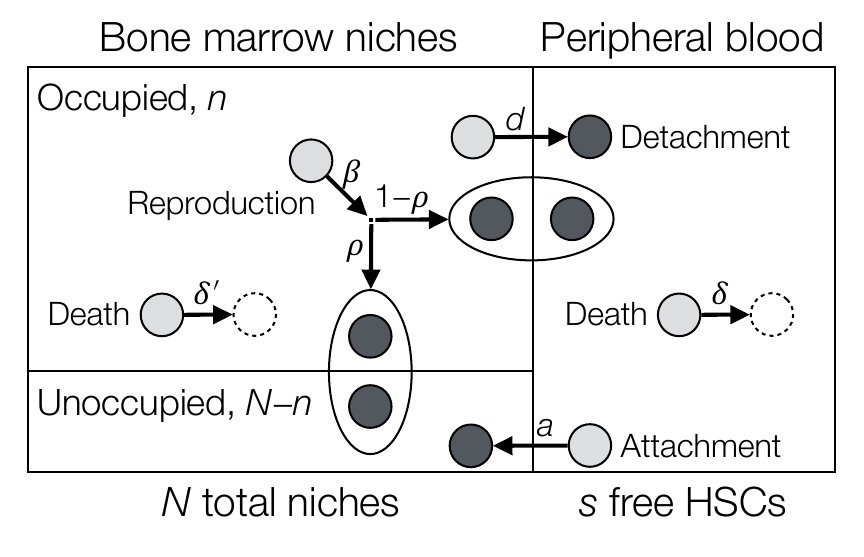}}

\caption{
Compartmental model for a single population of HSCs.
The bone marrow (BM) compartment has a fixed total of $N$ niches.
At a given time, $n$ of the niches are occupied, and $N-n$ remain unoccupied.
The peripheral blood (PB) compartment has no size restriction, and at a given time contains $s$ HSCs.
A HSC in the BM can detach at rate $d$ and enter the PB, while a cell in the PB can attach to an unoccupied niche with rate $a(N-n)/N$.
Here $(N-n)/N$ is the fraction of unoccupied niches.
HSCs may die in the PB or BM with rates $\delta$ and $\delta'$.
Reproduction (symmetric division) of HSCs occurs at rate $\beta$.
The new daughter cell attaches to an empty niche with probability $\rho$, otherwise it is ejected into the PB.
Dynamics are concretely described by the reactions in \eqref{eq:reactions}.
}
\label{fig:modelSchematic}
\end{figure}

Our model, shown schematically in \figref{fig:modelSchematic}, contains two compartments for the HSCs.
The bone marrow (BM) compartment consists in our model of a fixed number, $N$, of niches.
This means that a maximum of $N$ HSCs can be found there at any time, but generally the number of occupied niches is less than $N$.
The peripheral blood (PB) compartment, however, has no size restriction.
The number of cells in the PB and BM at a given time are given by $s$ and $n$, respectively.

The dynamics are indicated by arrows in \figref{fig:modelSchematic}.
Our model is stochastic and individual based, such that events are chosen randomly and waiting times between events are exponentially distributed.
Simulations are performed using the Gillespie stochastic simulation algorithm (SSA) \cite{gillespie:JPC:1977}.
HSCs in our model are only capable of dividing when attached to a niche; outside the niche, pre-malignant cells are incapable of proliferating due to the unfavourable conditions.
Upon division, the new daughter HSC enters another niche with probability $\rho$, or is ejected into the PB.
Here $\rho$ depends on the number of free niches, i.e. $\rho = \rho(n)$, and should satisfy $\rho(N)=0$, such that a daughter cell cannot attach if all niches are occupied.
Likewise, the migration of a HSC from the PB to the BM should depend on the number of empty niches.
We choose the attachment rate as $a(N-n)/N$ per cell.
In general, cells can die in both compartments.
However, we expect the death rate in the PB, $\delta$, to be higher than the death rate in the BM, $\delta'$, as the PB is a less favourable environment.

For our initial analysis, we assume there is no death in the BM compartment ($\delta'=0$), and new cells are always ejected into the PB ($\rho=0$).
These assumptions are relaxed in our detailed analysis, which can be found in the Supporting Information (SI).

A two-compartment model has been considered previously by Roeder and colleagues \cite{roeder:ExpHemat:2002,roeder:NM:2006}.
The rate of migration between the compartments is controlled by the number of cells in each compartment, as well as a cell-intrinsic continuous parameter which increases or decreases depending on which compartment the cell is in.
This parameter also controls the differentiation of the HSCs.
Further models of HSC dynamics, for example \cite{abkowitz:NatMed:1996,catlin:Blood:2005,dingli:CCY:2007,dingli:PLoSCB:2007,traulsen:Interface:2013}, have not considered the migration of cells between compartments.
For example, Dingli \emph{et al.} consider a constant-size population of HSCs in a homogeneous microenvironment \cite{dingli:CCY:2007,dingli:PLoSCB:2007}.
Competition between wildtype and malignant cells then follows a Moran process.
In our model the BM compartment size is fixed, but cell numbers can fluctuate.

To initially parametrise our model we consider only one species of HSCs: those which belong to the host.
In a steady-state organism, the number of HSCs in the PB and BM are close to their equilibrium values, which are labelled as $s^*$ and $n^*$, respectively.
These values have been reported previously in the literature for mice, and are provided in Table~\ref{tab:params}.
Other previously reported values include the total number of HSC niches $N$, HSC division rate $\beta$, and the time that cells spend in the PB, which we denote as $\ell$.
Using these values we can quantify the remaining model parameters $\delta$, $d$, and $a$.
These results are discussed in the next section.

\begin{table}
\centering
\caption{{\bf Parameter values from empirical murine observations. These are equilibrium values in healthy mice.}}
\label{tab:params}%
\begin{tabular}{|l|c|c|l|}
\hline
{\bf Description} & {\bf Parameter} & {\bf Value} & {\bf Reference}  \\ \hline
Total niches 				& $N$		& 10,000 niches	& \cite{bhattacharya:JEM:2006,bhattacharya:JEM:2009}	\\ \hline
Occupied niches 				& $n^*$		& 9,900 niches	& \cite{bhattacharya:JEM:2006,czechowicz:Science:2007}  \\ \hline
PB HSCs 						& $s^*$		& 1--100 cells	& \cite{bhattacharya:JEM:2009,wright:Science:2001}	\\ \hline
Average HSC division rate 	& $\beta$	& 1/39 per day	& \cite{takizawa:JEM:2011}	\\ \hline
Time in PB 					& $\ell$		& 1--5 minutes	& \cite{wright:Science:2001}	\\ \hline
\end{tabular}
\end{table}

When considering a second population of cells, such as a mutant clone or donor cells following transplantation, we may want to impose a selective effect relative to the host HSCs.
We therefore allow the mutant/donor cells to proliferate with rate $\beta_2 = (1 + \varepsilon) \beta$, where $\varepsilon$ represents the strength of selection.
For $\varepsilon = 0$, the mutant/donor cells proliferate at the same rate as the host HSCs.
In the SI we consider the general scenario of selection acting on all parameters.
Our analysis delivers an interesting result: the impact of selection on clonal expansion is independent of which parameter it acts on (provided $\delta'=\rho=0$).

For clonality and chimerism we use the same definition: the fraction of cells within the BM compartment that are derived from the initial mutant or the donor population of cells.
Typically, experimental measurements of clonality and chimerism use mature cells rather than HSCs.
However, this is beyond the scope of our model so we use HSC fraction as a proxy for this measurement.
We are therefore implicitly assuming that HSC chimerism correlates with mature cell chimerism.
The literature on the role of HSCs in native hematopoiesis is split \cite{sawai:Immunity:2016,sun:Nature:2014} (also reviewed in \cite{busch:CurrOpHem:2016}).
For the division rate of HSCs in mice we use the value $\beta = 1/39$ per day.
This is the average division rate of all HSCs within a host deduced from CFSE-staining experiments \cite{takizawa:JEM:2011}, but again there is some disagreement in reported values for this quantity \cite{wilson:Cell:2008,takizawa:JEM:2011,bernitz:Cell:2016}.
These differences arise from the interpretation of HSC cell-cycle dynamics.

More concretely, our model consists of four sub-populations: $n_1$ is the number of host or wildtype cells located in the BM, and $s_1$ is the number of cells of this type in the PB.
Likewise, $n_2$ and $s_2$ are the number of mutant/donor cells in the BM and PB, respectively.
The cell numbers are affected by the processes indicated in \figref{fig:modelSchematic} (with $\delta'=\rho=0$).
The effect of these events and the rate at which they happen are given by the following reactions:
\begin{linenomath}
\begin{subequations}
\label{eq:reactions}%
\begin{align}
\mbox{Reproduction:} \quad (n_i, s_i) &\xrightarrow{\makebox[7em]{$\beta_i n_i$}} (n_i, s_i+1), \\
\mbox{Death:} \quad (n_i, s_i) &\xrightarrow{\makebox[7em]{$\delta_i s_i$}} (n_i, s_i-1), \\
\mbox{Detachment:} \quad (n_i, s_i) &\xrightarrow{\makebox[7em]{$d_i n_i$}} (n_i-1, s_i+1), \\
\mbox{Attachment:} \quad (n_i, s_i) &\xrightarrow{\makebox[7em]{$a_i s_i (N-n)/N$}} (n_i+1, s_i-1),
\end{align}
\end{subequations}
\end{linenomath}
where $n = \sum_i n_i$, and $(N-n)/N$ is the fraction of unoccupied niches.
The corresponding deterministic dynamics are described by the ODEs:
\begin{linenomath}
\begin{subequations}
\label{eq:ODEsTwo}%
\begin{align}
\frac{\dd n_1}{\dd t} &= -d_1 n_1 + a_1 s_1 \frac{N-n}{N}, \\
\frac{\dd n_2}{\dd t} &= -d_2 n_2 + a_2 s_2 \frac{N-n}{N}, \\
\frac{\dd s_1}{\dd t} &= (d_1+\beta_1)n_1 - \left(\delta_1 + a_1 \frac{N-n}{N}\right)s_1, \\
\frac{\dd s_2}{\dd t} &= (d_2+\beta_2)n_2 - \left(\delta_2 + a_2 \frac{N-n}{N}\right)s_2.
\end{align}
\end{subequations}
\end{linenomath}
Recall we have $\beta_1 = \beta$ and $\beta_2 = (1+\varepsilon)\beta$ in the main manuscript, along with $\delta_1 = \delta_2 = \delta$, $a_1 = a_2 = a$, and $d_1 = d_2 = d$.

\subsection*{Accessibility}
A Wolfram Mathematica notebook containing the analytical details can be found at \url{https://github.com/ashcroftp/clonal-hematopoiesis-2017}.
This location also contains the Gillespie stochastic simulation code used to generate all data in this manuscript, along with the data files.

\section*{Results}
\subsection*{Steady-state HSC dynamics in mice}

By considering just the cells of the host organism, we can compute the steady state of our system from \eqref{eq:ODEsTwo}, and hence express the model parameters $\delta$, $d$, and $a$ in terms of the known quantities displayed in Table~\ref{tab:params}.
These expressions are shown in Table~\ref{tab:params2}, where we also enumerate the possible values of these deduced model parameters.
Even for the narrow range of values reported in the literature (Table~\ref{tab:params}), we find disparate dynamics in our model.
At one extreme, the average time a cell spends in the BM compartment ($1/d$) can be less than two hours (for $s^*=100$ cells and $\ell=1$ minute).
Thus under these parameters the HSCs migrate back-and-forth very frequently between the niches and blood, and the flux of cells between these compartments over a day ($s^*/\ell$) is significantly larger than the population size.
In fact, under these conditions $144,000$ HSCs per day leave the marrow and enter the blood.
With slower turnover in the PB compartment ($\ell = 5$ minutes, but still $s^* = 100$), the average BM residency time of a single HSC is eight hours, and $28,800$ HSCs leave the bone marrow per day.
At the other extreme, if the PB compartment is as small as reported in Ref.~\cite{bhattacharya:JEM:2009} ($s^*=1$ cell), then the residency time of each HSC in the bone marrow niche is between 8 and 290 days (for $\ell=1$ and $5$ minutes, respectively).
Under these conditions the number of cells entering the PB compartment per day is $1,440$ and $288$, respectively.
For an intermediate PB size of $s^* = 10$, the BM residency time is between $17$ and $90$ hours (for $\ell=1$ and $5$ minutes, respectively), and the flux of cells leaving the BM is a factor ten greater than for $s^* = 1$.

\begin{table}
\begin{adjustwidth}{-0.5in}{0in} 
\centering
\caption{{\bf Deduced model parameter values. The parameters $\delta$, $d$, and $a$ are given here as values per day. The remaining parameters ($N$, $\beta$, $n^*$) are given in Table~\ref{tab:params}}.}
\label{tab:params2}%
\begin{tabular}{|l|c|l|r c|c|c|c|c|c|}
\hline
				&			& 																		& 			& \multicolumn{6}{c}{{\bf Value (per day)}}  \\
{\bf Description} 		& {\bf Parameter}	& {\bf Expression} 															& $s^*$:		&\multicolumn{2}{c}{1 cell}	& \multicolumn{2}{c}{10 cells}	& \multicolumn{2}{c}{100 cells} \\
				& 			& 																		& $\ell$:	& 1 min 		& 5 mins			& 1 min		& 5 mins 			& 1 min		& 5 mins\\
\hline
Death rate 		& $\delta$ 	& $\frac{\beta n^*}{s^*}$ 												& 			& 250 		& 250 			& 25 		& 25 				& 2.5 		& 2.5 \\ \hline
\noalign{\smallskip}
Detachment rate 	& $d$ 		& $\frac{s^*}{\ell n^*} - \beta$ 										& 			& 0.12 		& 0.0034 		& 1.4 		& 0.27 				& 15 		& 2.9 \\ \hline
\noalign{\smallskip}
Attachment rate 	& $a$ 		& $\left(\frac{1}{\ell} - \frac{\beta n^*}{s^*}\right)\frac{N}{N-n^*}$	& 			& 120,000 	& 3,400 			& 140,000 	& 26,000 			& 140,000	& 29,000 \\ \hline
\end{tabular}
\end{adjustwidth}
\end{table}

\subsection*{Clonal dominance in mice}

Clonal dominance occurs when a single HSC generates a mature lineage which outweighs the lineages of other HSCs, or where one clone of HSCs outnumbers the others.
The definition of when a clone is dominant is not entirely conclusive.
Previous studies of human malignancies have used a variant allele frequency of $2\%$, corresponding to a clone that represents $4\%$ of the population \cite{steensma:Blood:2015,sperling:NatRevCancer:2017}.
For completeness we investigate clonality ranges from $0.1\%$ to $100\%$.

In the context of disease, this clone usually carries specific mutations which may confer a selective advantage over the wildtype cells in a defined cellular compartment.
The \emph{de novo} emergence of such a mutant occurs following a reproduction event.
Therefore, in our model with $\rho=0$, after the mutant cell is generated it is located in the PB compartment, and for the clone to expand it must first migrate back to the BM.
This initial phase of the dynamics is considered in general in the next section of transplant dynamics, where a positive number $\mathcal{S}$ of mutant/donor cells are placed in the PB.
We find (as shown in the SI) that the expected number of these cells that attach to the BM after this initial dynamical phase is
\begin{equation}
n_2
= \frac{a(N-n^*)/N}{\delta + a(N-n^*)/N}\mathcal{S}
= \left(1-\frac{\beta \ell n^*}{s^*}\right)\mathcal{S}.
\label{eq:chimerismLow}
\end{equation}
We then apply a fast-variable elimination technique to calculate how long it takes for this clone to expand within the host \cite{constable:PRE:2014,constable:PRL:2015}.
This procedure reduces the dimensionality of our system, and makes it analytically tractable.
A full description of the analysis can be found in the SI, but we outline the main steps and results of this procedure below.

We first move from the master equation -- the exact probabilistic description of the stochastic dynamics -- to a set of four stochastic differential equations (SDEs) for each of the variables via an expansion in powers of the large parameter $N$ \cite{gardiner:book:2009}.
We then use the projection method of Constable \emph{et al.} \cite{constable:PRE:2014,constable:PRL:2015} to reduce this system to a single SDE describing the relative size of the clone.
This projection relies on the weak-selection assumption, i.e. $0 \le \varepsilon \ll 1$.
The standard results of Brownian motion are then applied to obtain the statistics of the clone's expansion.
In particular, the probability that the mutant/donor HSCs reach a fraction $0 < \sigma \le 1$ of the occupied BM niches is given by
\begin{equation}
\phi(z_0, \sigma) = \frac{1 - e^{-\Lambda z_0}}{1 - e^{-\Lambda \sigma \xi}},
\label{eq:selectiveDominanceProb}
\end{equation}
where $z_0$ is the initial clone size can be found explicitly from \eqref{eq:chimerismLow}, such that $z_0=n_2/N$.
We also have $\xi = n^*/N$, and $\Lambda$ is a constant describing the strength of deterministic drift relative to stochastic diffusion.
Concretely, we have
\begin{equation}
\Lambda
= \varepsilon N \frac{d\beta + d\delta + \beta\delta}{(d+\beta)\delta}
= \varepsilon N \left(1 + \frac{s^*}{n^*} - \beta \ell \right).
\end{equation}
The mean time for the clone to expand to size $\sigma$ (i.e. the mean conditional time) is written as $T_\xi(z_0, \sigma) = \theta(z_0, \sigma) / \phi(z_0, \sigma)$, where $\theta(z_0, \sigma)$ is given by the solution of
\begin{equation}
\frac{\partial^2 \theta(z_0, \sigma)}{\partial z_0^2} + \Lambda \frac{\partial \theta(z_0, \sigma)}{\partial z_0} = -\frac{N}{\mathcal{B}} \frac{\phi(z_0, \sigma)}{z_0(\xi-z_0)}, \quad \theta(0) = \theta(\sigma \xi) = 0. 
\label{eq:selectiveDominanceTime}
\end{equation}
Here $\mathcal{B}$ is another constant describing the magnitude of the diffusion, and is given by
\begin{equation}
\mathcal{B}
= \frac{d(d+\beta)\beta\delta^2}{\xi(d\beta + d\delta + \beta\delta)^2}
= \frac{\beta N}{s^*} \frac{\frac{s^*}{n^*} - \beta \ell}{\left(1+\frac{s^*}{n^*}-\beta\ell\right)^2}.
\end{equation}
Although a general closed-form solution to \eqref{eq:selectiveDominanceTime} is possible, it is too long to display here.
Instead we use an algebraic software package to solve the second-order differential equation.
A similar expression to \eqref{eq:selectiveDominanceTime} can be obtained for the second moment of the fixation time, as shown in \cite{goel:book:1974} and repeated in the SI.

\begin{figure}[h]
\centering

\iftoggle{showFigs}{\includegraphics[width=\textwidth]{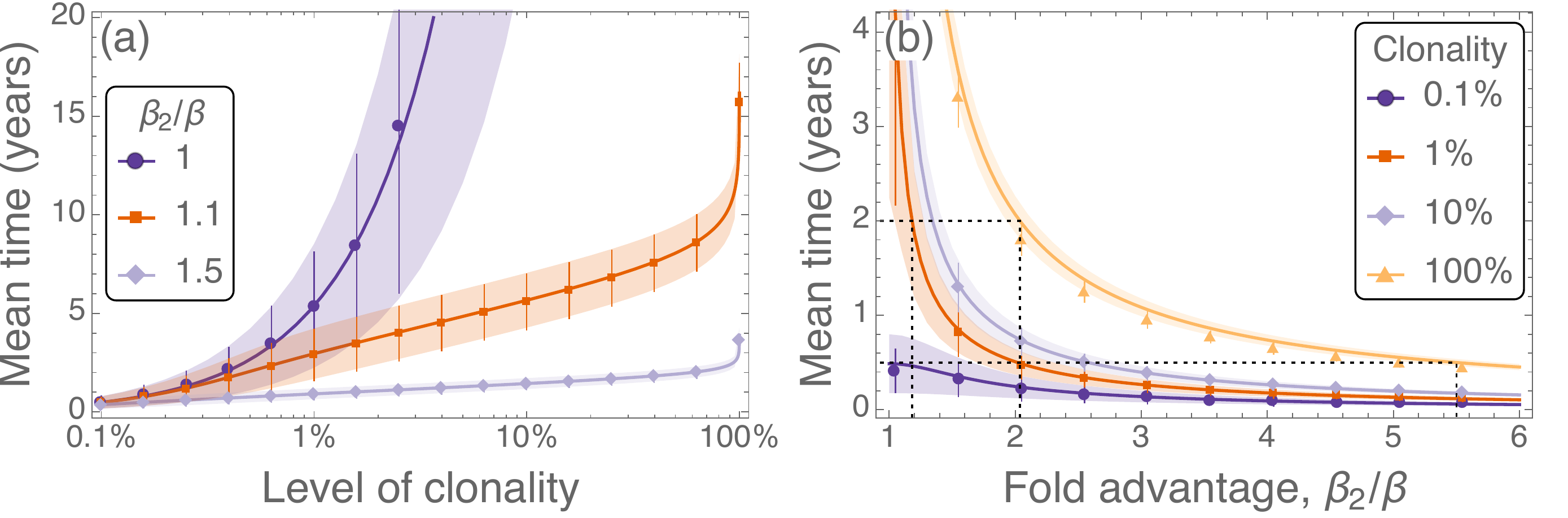}}

\caption{
Time taken for a clone initiated from a single HSC to expand under different levels of selection.
(a) Time taken for a mutant clone to expand as a function of the level of clonality reached, with colour indicating the selective effect of the mutant.
(b) Time taken for a mutant clone to expand as a function of the selective effect, with colour indicating different levels of clonality.
Symbols are results from $10^3$ simulations of the full model (with associated standard deviations), and solid lines are predictions from \eqref{eq:selectiveDominanceTime}.
Shaded regions are the predicted standard deviations, using the formula presented in the SI.
Here $\ell = 3$ minutes, $s^* = 100$, and the remaining parameters are as in Table~\ref{tab:params}.
}
\label{fig:dominanceMouse}
\end{figure}

The first scenario we consider is the expansion of a neutral clone ($\varepsilon = 0$); i.e. how likely is it that a single cell expands into a detectable clone in the absence of selection?
It is known that the time to fixation of a neutral clone in a fixed-size population grows linearly in the system size \cite{kimura:Genetics:1969}.
Interestingly and importantly, in intestinal crypts this fixation is seen frequently because $N = \mathcal{O}(10)$ \cite{snippert:cell:2010}.
In the hematopoietic system, however, it likely takes considerably longer than this due to the relatively large number of stem cells.
Solving \eqref{eq:selectiveDominanceTime} with $\varepsilon = 0$ gives the mean conditional expansion time as
\begin{equation}
T_\xi(z_0, \sigma) = \frac{N}{\mathcal{B}} \left[ \frac{\xi-z_0}{z_0}\log\left(\frac{\xi}{\xi-z_0}\right) + \frac{1-\sigma}{\sigma} \log(1-\sigma)\right].
\end{equation}
From this solution we find that it takes, on average, $5$--$45$ years for a neutral clone to reach $1\%$ clonality ($\sim 100$ HSCs).
Expanding to larger sizes takes considerably longer, as highlighted in \figref{fig:dominanceMouse}.
Therefore, clonal hematopoiesis in mice is unlikely to result from neutral clonal expansion; for a clone to expand within the lifetime of a mouse it must have a selective advantage.
Neutral results for human systems are considered in the discussion.

When the mutant clone has an advantage, there is always some selective force promoting this cell type.
Therefore the probability of such a clone expanding is higher than the neutral case, as seen from \eqref{eq:selectiveDominanceProb}.
In \figref{fig:dominanceMouse} we illustrate the time taken for a single mutant HSC to reach specified levels of clonal dominance for different selective advantages.
Advantageous clones ($\beta_2/\beta > 1$) initially grow exponentially in time [\figref{fig:dominanceMouse}(a)], and are much faster than neutral expansion ($\beta_2/\beta=1$).
These clones can reach levels of up to $90\%$ relatively quickly, however replacing the final few host cells takes much longer.
The advantage that a mutant clone must have if it is to represent a certain fraction of the population in a given period of time can be found from \figref{fig:dominanceMouse}(b).
For a single mutant to completely take over in two years, it requires a fold reproductive advantage of $\beta_2/\beta \approx 2$ [dashed lines in \figref{fig:dominanceMouse}(b)].
This means that the cells in this clone are dividing at least twice as fast as the wildtype host cells.
To achieve $1\%$ clonality in this timeframe, the advantage only has to be $\beta_2/\beta \approx 1.2$.
For the clone to expand in shorter time intervals, a substantially larger selective advantage is required.
For example, $100\%$ clonality in six months from emergence of the mutant requires $\beta_2/\beta \approx 5.5$, i.e. the dominant clone needs to divide more than five times faster than the wildtype counterparts.

As shown in the SI, \eqref{eq:selectiveDominanceProb} and \eqref{eq:selectiveDominanceTime} are equivalent to the results obtained from a two-species Moran process.
This suggests the two-compartment structure is not necessary to capture the behaviour of clonal dominance.
However, the consideration of multiple compartments is required to understand transplantation dynamics, as covered in the next section.

\subsection*{Transplant success in mice}

We now turn our attention to the scenario of HSC transplantation.
As previously mentioned this situation is analogous to the disease spread case, with the exception that the initial `dose' of HSCs can be larger than one.
We first consider the case of a non-preconditioned host.
We then move onto transplantation in preconditioned hosts, where all host cells have been removed.

\subsubsection*{Engraftment in a non-preconditioned host}

Multiple experiments have tested the hypothesis that donor HSCs can engraft into a host which has not been pretreated to remove some or all of the host organism's HSCs \cite{stewart:Blood:1993,quesenberry:Blood:1994,rao:ExpHemat:1997,blomberg:ExpHemat:1998,slavin:Blood:1998,quesenberry:NYAS:2001,bhattacharya:JEM:2006,bhattacharya:JEM:2009,takizawa:JEM:2011,kovtonyuk:Blood:2016}.
These studies have found that engraftment can be successful; following repeated transplantations mice display a chimerism with up to $40\%$ of the HSCs deriving from the donor \cite{quesenberry:Blood:1994,rao:ExpHemat:1997,blomberg:ExpHemat:1998}.

In this scenario we start with a healthy host organism and inject a dose of $\mathcal{S}$ donor HSCs into the PB compartment, in line with the experimental protocols mentioned above.
These donor cells can be neutral, or may have a selective (dis)advantage.
Injecting neutral cells reflects the \emph{in vivo} experiments described above, while advantageous cells can be used to improve the chances of eliminating the host cells.
Transplanting disadvantageous cells would reflect the introduction of `normal' HSCs into an already diseased host carrying advantageous cells.
We do not consider this scenario further here, as the diseased cells are highly unlikely to be replaced without host preconditioning.

We can separate the engraftment dynamics of these donor cells into two different regimes: i) the initial relaxation to a steady state where the total number of HSCs is stable, and ii) long-time dynamics eventually leading to the extinction of either the host or donor HSCs.
We focus on these regimes separately.
Upon the initial injection of the donor HSCs, the PB compartment contains more cells than the equilibrium value $s^*$.
This leads to a net flux of cells attaching to the unoccupied niches in the BM until the population relaxes to its equilibrium size.
Once the equilibrium is reached, the initial dynamics end, and the long-term noise-driven dynamics take over (discussed below).
The challenge for this first part is to determine how many of the donor HSCs have attached to the BM at the end of the initial dynamics.

We identify two distinct behaviours which occur under low and high doses of donor HSCs.
If the dose is small ($\mathcal{S} \ll N-n^*$), then the number of donor HSCs that attach to the BM is given by \eqref{eq:chimerismLow}, and is proportional to the dose size $\mathcal{S}$.
To obtain this result we have assumed that the number of occupied niches remains constant, such that each donor cell has the same chance of finding an empty niche.
However, if the dose of donor HSCs is large enough then all niches become occupied and the BM compartment is saturated; attachment to the BM can only occur following a detachment.
Using this assumption, the initial dynamics can then be described by the linear ODEs
\begin{linenomath}
\begin{subequations}
\label{eq:chimerismHigh}%
\begin{align}
\frac{\dd n_2}{\dd t} &= -d n_2 +\frac{dN}{s(t)}s_2, \\
\frac{\dd s_2}{\dd t} &= (\beta+d)n_2 - \left(\delta + \frac{dN}{s(t)}\right)s_2,
\end{align}
\end{subequations}
\end{linenomath}
where $s(t)$, the total number of cells in the PB compartment, is found from $\dot{s} = \beta N - \delta s$.
A derivation of \eqref{eq:chimerismHigh} can be found in the SI.

\begin{figure}[h]
\centering

\iftoggle{showFigs}{\includegraphics[width=\textwidth]{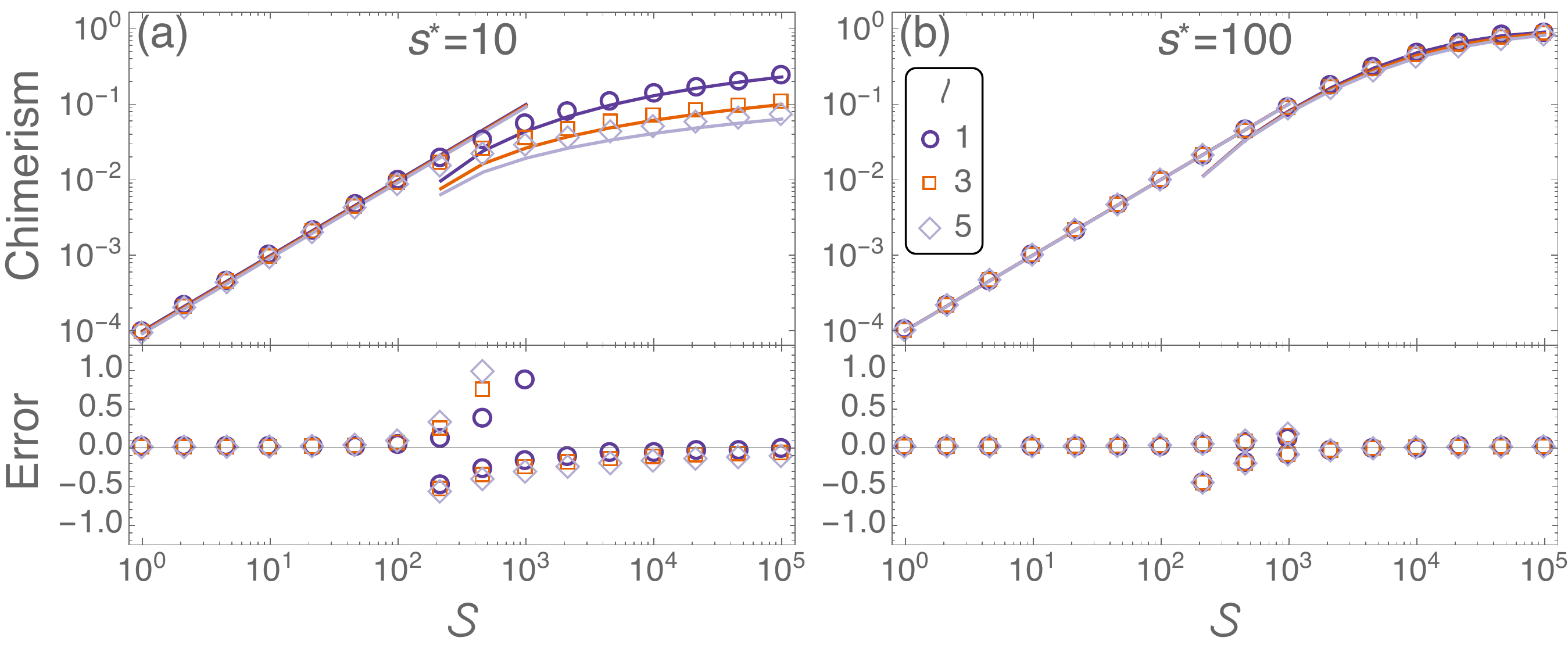} }

\caption{
Initial chimerism of neutral donor cells in a healthy, non-preconditioned host.
Upper panels depict the level of donor chimerism shortly after a dose of neutral donor cells, $\mathcal{S}$, is injected into the host.
Symbols are from numerical integration of \eqref{eq:ODEsTwo}.
The small-dose regime is described by \eqref{eq:chimerismLow} (solid lines for $\mathcal{S} < 10^3$), and the large-dose regime is described by \eqref{eq:chimerismHigh} (solid lines for $\mathcal{S} > 10^2$).
Lower panels show the accuracy of these approximations when compared to the numerical integration of \eqref{eq:ODEsTwo}.
This error takes the form (\emph{approx.}$-$\emph{exact})/\emph{exact}.
(a) $s^* = 10$, and (b) $s^* = 100$.
The lifetime in the PB, $\ell$, is measured in minutes.
Remaining parameters are as in Table~\ref{tab:params}.
}
\label{fig:initialChimerism}
\end{figure}

The predicted chimerism, and the accuracy of these predictions, at the end of the initial phase are shown in \figref{fig:initialChimerism}.
The efficiency of donor cell engraftment decreases in the large-dose regime ($\mathcal{S} > N-n^*$).
This is simply because the niche-space is saturated, so HSCs spend longer in the blood and are more likely to perish.
If the lifetime in the PB ($\ell$) is short, then we have more frequent migration between compartments, as highlighted in Table~\ref{tab:params2}.
Hence smaller $\ell$ leads to higher chimerism. 
The approximation from \eqref{eq:chimerismHigh} becomes increasingly accurate for larger doses.
The two approximations break down at the cross-over region between small and large doses.
In this regime the number of occupied niches does not reach a stable value.

If the donor cells have a selective (dis)advantage, then the deterministic dynamics predict the eventual extinction of either the host or donor cells.
However, the selective effect is usually small and only acts on a longer timescale.
Therefore the initial dynamics are largely unaffected by selection, and we assume neutral donor cell properties when we model the initial dynamics.

\begin{figure}[h]
\centering

\iftoggle{showFigs}{\includegraphics[width=0.5\textwidth]{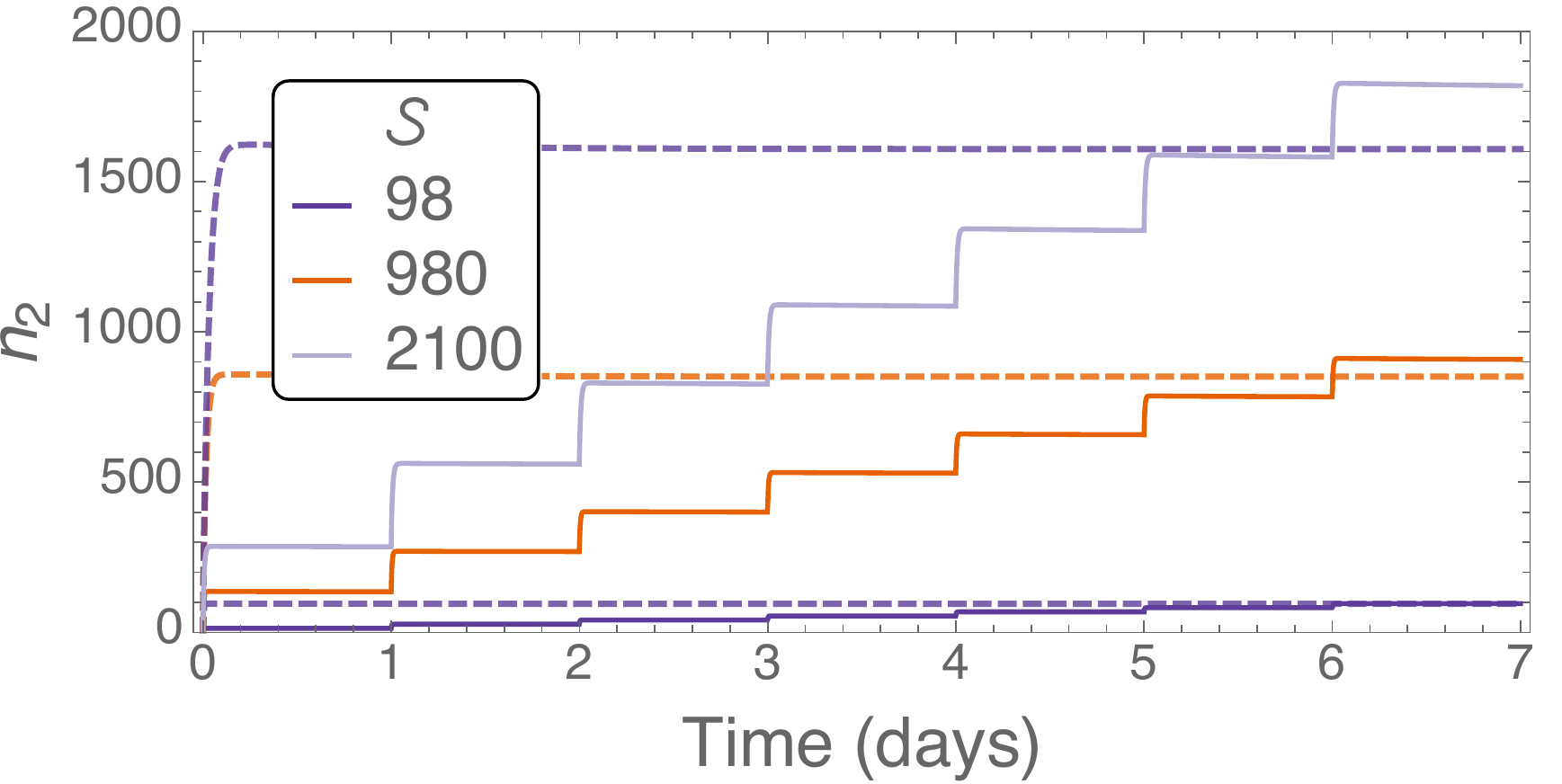}}

\caption{
Number of donor HSCs attaching to the BM of a non-preconditioned host after a single dose (dashed lines) or seven daily doses (solid lines).
Both treatments use the same total number, $\mathcal{S}$, of donor HSCs.
Trajectories are from numerical integration of the ODEs~\eqref{eq:ODEsTwo}.
Here we have $\ell=3$ minutes, $s^*=100$, and the remaining parameters are as in Table~\ref{tab:params}.
}
\label{fig:multipleDoses}
\end{figure}

The inefficiency of large doses can be overcome by administering multiple small doses over a long period.
In this way we prevent the niches from becoming saturated and fewer donor cells die in the PB.
Hence we should be able to obtain a higher level of engraftment when compared to a single-bolus injection of the same total number of donor HSCs.
These effects have been tested experimentally \cite{quesenberry:Blood:1994,rao:ExpHemat:1997,blomberg:ExpHemat:1998,bhattacharya:JEM:2009}.
Parabiosis experiments are also an extreme example of this; they represent a continuous supply of donor cells \cite{wright:Science:2001}. 
As shown in \figref{fig:multipleDoses}, our model captures the same qualitative behaviour as reported in the experiments: Multiple doses lead to higher levels of chimerism at the end of the initial phase of dynamics.
This effect is highlighted more when the total dose size is large.
Using our analysis we know how efficient each dose is, and what levels of chimerism can be achieved.
Hence our model can be used to optimise dosing schedules such that they are maximally efficient.

We note here that engraftment efficiency is only important when donor cells are rare and there is no danger to life.
This is the case, for example, in experimental protocols when tracking small numbers of cells.
It makes sense to here use the multiple dosing strategy.
Transplantation following preconditioning, however, provides a more viable approach to disease treatment where patient survival needs to be maximised.
In this case the dose size should be increased, but it should be considered that there are diminishing returns in engraftment when the dose size is large enough to saturate all open niches.
This dose size can be read from \figref{fig:initialChimerism}.

The long-term dynamics are handled in the same way as the clonal dominance results above; we use \eqref{eq:selectiveDominanceProb} and \eqref{eq:selectiveDominanceTime} to show how the number of donor cells injected into the PB affects the probability that they expand (as opposed to die out), and the time it takes for the host cells to be completely displaced.
One key result is that a dose of just eight HSCs with an advantage of $\beta_2/\beta = 1.1$ has over $50\%$ chance to fixate in the host.
However, the time for this to happen is $\sim 16$ years.
With a reproductive advantage of $\beta_2/\beta = 1.5$ the success rate is $\sim 95\%$ for the same dose, and the time taken now falls to $\sim 4$ years.
Further results are found in the SI.

\subsubsection*{Engraftment in a preconditioned host}

HSC transplantation procedures are often preceded by treatment or irradiation of the host -- referred to as host preconditioning.
This greatly reduces the number of host HSCs in the BM compartment.
For this section we assume complete conditioning such that no host HSCs remain, i.e. myeloablative conditioning.
Following the pre-treatment, a dose of donor HSCs of size $\mathcal{S}$ is injected into the PB compartment.
We then want to know the probability that these cells reconstitute the organism's hematopoietic system.
For this section we assume that donor HSCs have identical properties to the wildtype cells (i.e. no selection).
We further assume that all donor HSCs have the potential to reconstitute the hematopoietic system in the long-term --
in experiments this is not always the case as not all cells which are sorted as phenotypic HSCs (as defined by surface markers) are functional, reconstituting HSCs (see e.g. \cite{matsuzaki:Immunity:2004}).
Because of this assumption, we only show results for the injection of a single donor HSC into the conditioned host ($\mathcal{S} = 1$).
Higher doses lead to a greater probability of reconstitution.
A further assumption is, that the host maintains (or is provided with) enough mature blood cells during the reconstitution period to sustain life.

\begin{figure}[h]
\centering

\iftoggle{showFigs}{\includegraphics[width=0.5\textwidth]{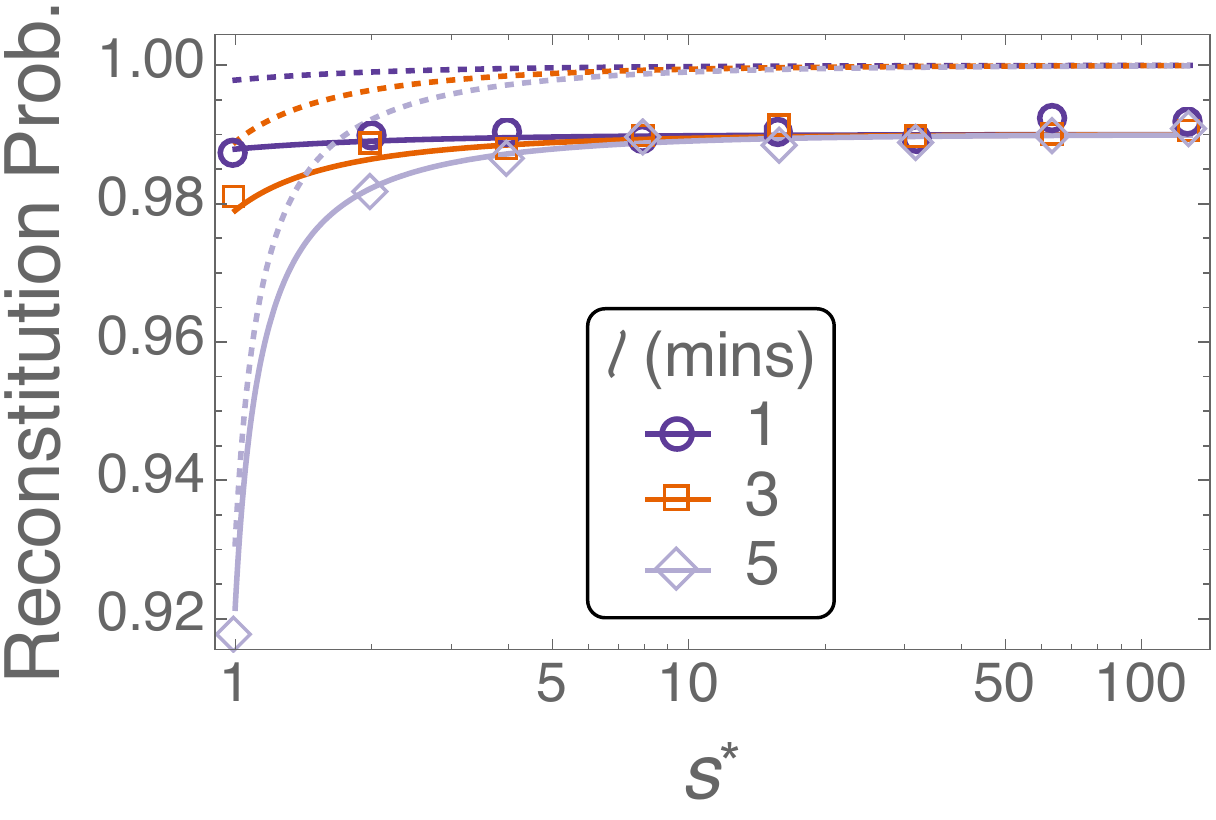}}

\caption{
Probability of reconstitution from a single donor HSC which is injected into a preconditioned host.
Symbols are results from $10^5$ simulations of the stochastic model.
For efficiency we ran the simulations until the population reached either 0 (extinction) or 100 (reconstitution), and we assume no further extinction events occur once this upper limit has been reached.
Dotted lines are the `first-order' prediction of \eqref{eq:emptyReconst0}.
Solid lines are the predictions of \eqref{eq:emptyReconst} which account for detachments, reattachments, and reproduction events.
Remaining parameters are as in Table~\ref{tab:params}.
}
\label{fig:conditionedReconstitution}
\end{figure}

We consider two approaches for estimating the probability of hematopoietic reconstitution.
As a first-order approximation, the probability that a single HSC in the PB compartment dies is $\psi = \delta/(\delta+a)$.
Here we have assumed that all niches are unoccupied, such that the attachment rate per cell is $a(N-0)/N=a$.
For a dose of size $\mathcal{S}$, the reconstitution probability is $\varphi = 1-\psi^{\mathcal{S}}$.
Hence we have
\begin{equation}
\varphi = 1 - \left(\frac{\delta}{\delta + a}\right)^\mathcal{S}.
\label{eq:emptyReconst0}
\end{equation}
This prediction, \eqref{eq:emptyReconst0}, is shown as dotted lines in \figref{fig:conditionedReconstitution}, which, however, does not agree with results from the model.
The second approach considers all possible combinations of detachments and reattachments, as well as reproduction events.
This leads to a reconstitution probability, given a dose of $\mathcal{S}$ donor cells, of
\begin{equation}
\varphi = 1 - \left(\frac{\delta}{\delta + a} \frac{d + \beta}{\beta}\right)^{\mathcal{S}},
\label{eq:emptyReconst}
\end{equation}
which is derived in the SI.
This result, \eqref{eq:emptyReconst}, is shown as solid lines in \figref{fig:conditionedReconstitution}, and is in excellent agreement with the reconstitution probability observed in simulations.
From these results we can conclude that, in our model, HSCs migrate multiple times between the PB and BM before they establish a sustainable population.
It is also the case that in this model a single donor HSC is sufficient to repopulate a conditioned host in $\sim 90$ -- $99\%$ of cases across all the parameter ranges reported in Table~\ref{tab:params}.

\section*{Discussion}
We have introduced a mathematical model that describes the back-and-forth migration of hematopoietic stem cells between the blood and bone marrow within a host.
This is motivated by the literature of HSC dynamics in mice.
The complexity of the model has been kept to a minimum to allow us to parametrise it based on empirical results.
The model is also analytically tractable, permitting a more thorough understanding of the dynamics and outcomes.
For example, on long timescales we find the that the two-compartment model is equivalent to the well-studied Moran model.
Meanwhile, analysis of the reconstitution of a preconditioned mouse shows that in our model HSCs migrate multiple times between
the BM and PB compartments before establishing a sustainable population.

Given these dynamics we first investigate clonal dominance, where a clone originating from a single mutant cell expands in the HSC population.
In mice we find that a selective advantage is required if the clone is to be detected within a lifetime:
A clone starting from a single cell with a reproduction rate $50\%$ higher than the wildtype can expand to $1\%$ clonality in one year.
A cell dividing twice as fast as the wildtype reaches ${>10\%}$ clonality in the same timeframe.
Such division rates can be reached by MPN-initiating HSCs \cite{lundberg:JEM:2014}.
The requirement of a selective advantage agrees with the clinical literature where, for example, mutants are known to enjoy a growth advantage under inflammatory conditions \cite{fleischman:Blood:2011,kleppe:CanDisc:2015}.

The model also captures the scenario of stem cell transplantation.
Engraftment into a non-preconditioned host is analogous to clonal dominance, except that the clone is initiated by multiple donor cells.
For small doses of donor HSCs, the number of cells that attach to the BM is directly proportional to the size of the dose.
For larger doses the BM niches are saturated, leading to lower engraftment efficiency.
Donor chimerism can be improved by injecting the host with multiple small doses, as opposed to a large single-bolus dose of the same size.
This agrees with results that have been reported in the empirical literature \cite{quesenberry:Blood:1994,rao:ExpHemat:1997,blomberg:ExpHemat:1998,bhattacharya:JEM:2009}.
Following preconditioning of a mouse to remove all host cells, we find that a single donor HSC is sufficient to repopulate a host in ${\sim \! 90}$--$99\%$ of cases.
This result rests on the assumption that the donor stem cell was, in fact, a long-term reconstituting HSC, which may not be the case in experimental setups.

In the SI we consider the effects of death occurring in the BM niche ($\delta' \ne 0$), and the direct attachment of a new daughter cell to the bone marrow niche ($\rho \ne 0$).
We find that death in the niche increases the migration rate of cells between the PB and BM compartments, which can greatly reduce the attachment success of the low-frequency mutant/donor cells.
However, the direct attachment of daughter cells to the niche has no effect on the initial attachment of donor/mutant cells, and on the level of chimerism achieved in the initial phase of the dynamics.

\begin{figure}[h]
\centering

\iftoggle{showFigs}{\includegraphics[width=0.5\textwidth]{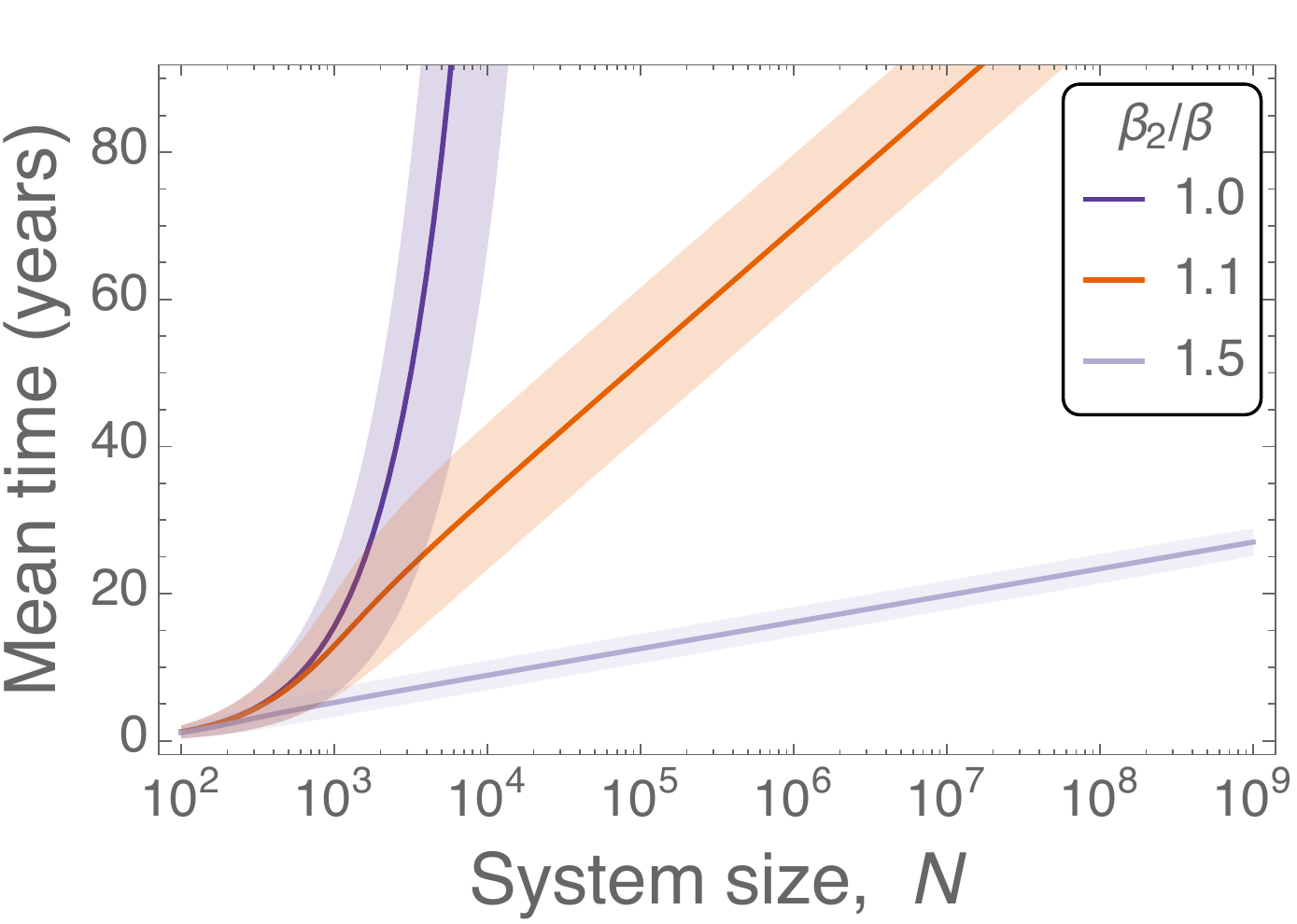}}

\caption{
Time taken until a clone initiated from a single cell represents $4\%$ \cite{steensma:Blood:2015,sperling:NatRevCancer:2017} of the human HSC pool, as a function of the total number of niches in the system.
Colours represent the selective advantage of the invading clone.
Lines are given by the solution of \eqref{eq:selectiveDominanceTime}, and shaded regions represent the calculated standard deviation (details in the SI).
Remaining parameters are $\beta=1/40$ week$^{-1}$ \cite{catlin:Blood:2011}, $\ell=60$ minutes, $s^*=0.01N$, and $n^*=0.99N$.
Here $\ell$, $s^*$, and $n^*$ are extrapolated from the murine data, where $\ell_{\rm human} \approx 10 \ell_{\rm mouse}$, which follows the same scaling as the HSC division rate, $\beta$.
Further parameter combinations are shown in Fig~S1 and Fig~S2.
References refer only to the source of parameters; no part of this figure has been reproduced from previous works.
}
\label{fig:dominanceMan}
\end{figure}

Broadening the scope of our investigation, clonality of the hematopoietic system is a major concern for human health \cite{genovese:NEJM:2014,jaiswal:NEJM:2014,xie:NatMed:2014,steensma:Blood:2015}.
Clinical studies have shown that $10\%$ of people over $65$ years of age display clonality, yet $42\%$ of those developing hematologic cancer displayed clonality prior to diagnosis \cite{genovese:NEJM:2014}.
Our model, and the subsequent analysis, can be applied to this scenario.
However, the number of HSCs in man is debated, with estimates of ${\sim \! 400}$ \cite{buescher:JCI:1985,dingli:PLoSONE:2006}, 
$\mathcal{O}(10^4)$ \cite{abkowitz:Blood:2002}, or $\mathcal{O}(10^7)$ \cite{nombela:BloodAdv:2017}.
Estimates as high as $\mathcal{O}(10^9)$ can also be obtained by combining the total number of nucleated bone marrow cells \cite{harrison:JCP:1962} with stem cell fraction measurements \cite{wang:Blood:1997,kondo:ARI:2003,pang:PNAS:2011}.
In \figref{fig:dominanceMan} we summarise how neutral and advantageous clones starting from a single HSC expand in human hematopoietic systems.
We find that $4\%$ clonality \cite{steensma:Blood:2015,sperling:NatRevCancer:2017} can be achieved in a short period of time for even neutral clones [\figref{fig:dominanceMan}(a)].
If the human HSC pool is $\mathcal{O}(10^3)$ or smaller, we would expect clonal hematopoiesis and the associated malignancies to be highly abundant in the population, perhaps more-so than they currently are \cite{genovese:NEJM:2014,jaiswal:NEJM:2014,xie:NatMed:2014,steensma:Blood:2015}.
On the other hand, for a system size of $N = 10^6$ it takes thousands of years for a single neutral HSC to expand to detectable levels, making neutral expansion extremely unlikely to result in clonal hematopoiesis.
Therefore, for clonal hematopoiesis to occur in a pool of this size or larger \cite{nombela:BloodAdv:2017} the mutants would require a significant fitness advantage over the wildtype HSCs.
We also consider a range of parameters, and even relax the $\alpha=\varrho=0$ condition, in Fig~S1 and Fig~S2.
We find no significant differences in the predictions of our model.

\subsection*{Limitations}
Our model has been kept to a minimal level of biological detail to allow for parametrisation from experimental results.
This has the added benefit of analytic tractability.
The model is constructed under steady-state conditions, which is the case for neutral clonal expansion.
However, in the case of donor-cell transplantation following myeloablative preconditioning, we are no longer in a steady state.
Here we expect some regulatory mechanisms to affect the HSC dynamics, including a faster reproductive rate and a reduced probability of cells detaching from the niche.
There are also possibilities for mutants to exploit or evade the homeostatic mechanisms \cite{brenes:PNAS:2011}.
Different mechanisms of stem cell control have recenty been considered for hematopoietic cells \cite{stiehl:BMT:2014}, as well as in colonic crypts \cite{yang:JTB:2017}.

The steady state assumption is also unable to capture the different dynamics associated with ageing.
For example, in young individuals the hematopoietic system is undergoing expansion.
In our model there is no distinction between young and old systems.
In Fig~S3 we demonstrate the impact of a (logistically) growing number of niches.
Such growth means clonal hematopoiesis is likely to be detected earlier, and therefore would increase our lower bound estimate on the number of HSCs in man.
Telomere-length distributions have been used to infer the HSC dynamics from adolescence to adulthood, and have suggested a slowing down of HSC divisions as life progresses \cite{werner:eLife:2015}.
Faster dynamics in early life would lead to a higher incidence among young people, which again increases our lower bound estimate.

It is also not entirely clear how to extrapolate the parameters from the reported mouse data to a human system.
Here we have taken the simplest approach and appropriately scaled the unknown parameters.
However, hematopietic behaviour may differ between species.
For example, results of HSC transplantation following myeloablative therapy in non-human primates have shown that clones of hematopoietic cells persist for many years \cite{kim:Blood:2009,kim:CSC:2014}.
This could be due to single HSCs remaining attached to the niche and over-contributing to the hematopoietic system, or due to clonal expansion of the HSCs to large enough numbers such that a contributing fraction will always be found in the BM.
Both of these mechanisms are features of our model: the time a cell spends in the BM is much longer than the time in the PB and can be increased further by tuning the model parameters, namely by decreasing $s^*$ or increasing $\ell$.
Changes to these parameters seems to have little effect on our predictions of clonal expansion, as shown in Fig~S1 and Fig~S2.
Clonal extinctions are also a feature of our work, and have been identified in non-human primates \cite{kim:CSC:2014}.

A more general point to discuss is the role of hematopoietic stem cells in blood production.
In our model we are only considering HSC dynamics, however it has been proposed that downstream progenitor cells are responsible for maintaining hematopoiesis \cite{schoedel:Blood:2016} in mice.
Hence, myeloid clonality would also be determined by the behaviour of these progenitor cells.
On the other hand, an independent study found that HSCs are driving multi-lineage hematopoiesis \cite{sawai:Immunity:2016}, suggesting we are correct in our approach.
Again we also expect there to be variation between species in this balance of HSC/progenitor activity.
With little quantitative information available, we have assumed that HSCs are the driving force of steady-state hematopoiesis across mice and humans.

\subsection*{Conclusion}
In conclusion, this simple mathematical model encompasses multiple HSC-engraftment scenarios and qualitatively captures empirically observed effects.
The mathematical calculations provide insight into how the dynamics of the model unfold.
The analytical results, which we have verified against stochastic simulations, allow us to easily investigate how parameter variation affects the outcome.
We now hope to extend this analysis, incorporating further effects of disease and combining this model with the differentiation tree of hematopoietic cells.

\section*{Supporting Information}

\paragraph*{SI}
\label{SI}
{\bf Supporting mathematical details.}
Contains detailed derivations of all equations presented in the manuscript, including the details of the projection method.
The analysis is carried out for unrestricted parameters, including selective effects on all parameters and permitting death in the BM space, as well as direct attachment of new daughter cells to the niche.

\iftoggle{showSuppFigs}{\begin{figure*}[h]\centering\includegraphics[width=\textwidth]{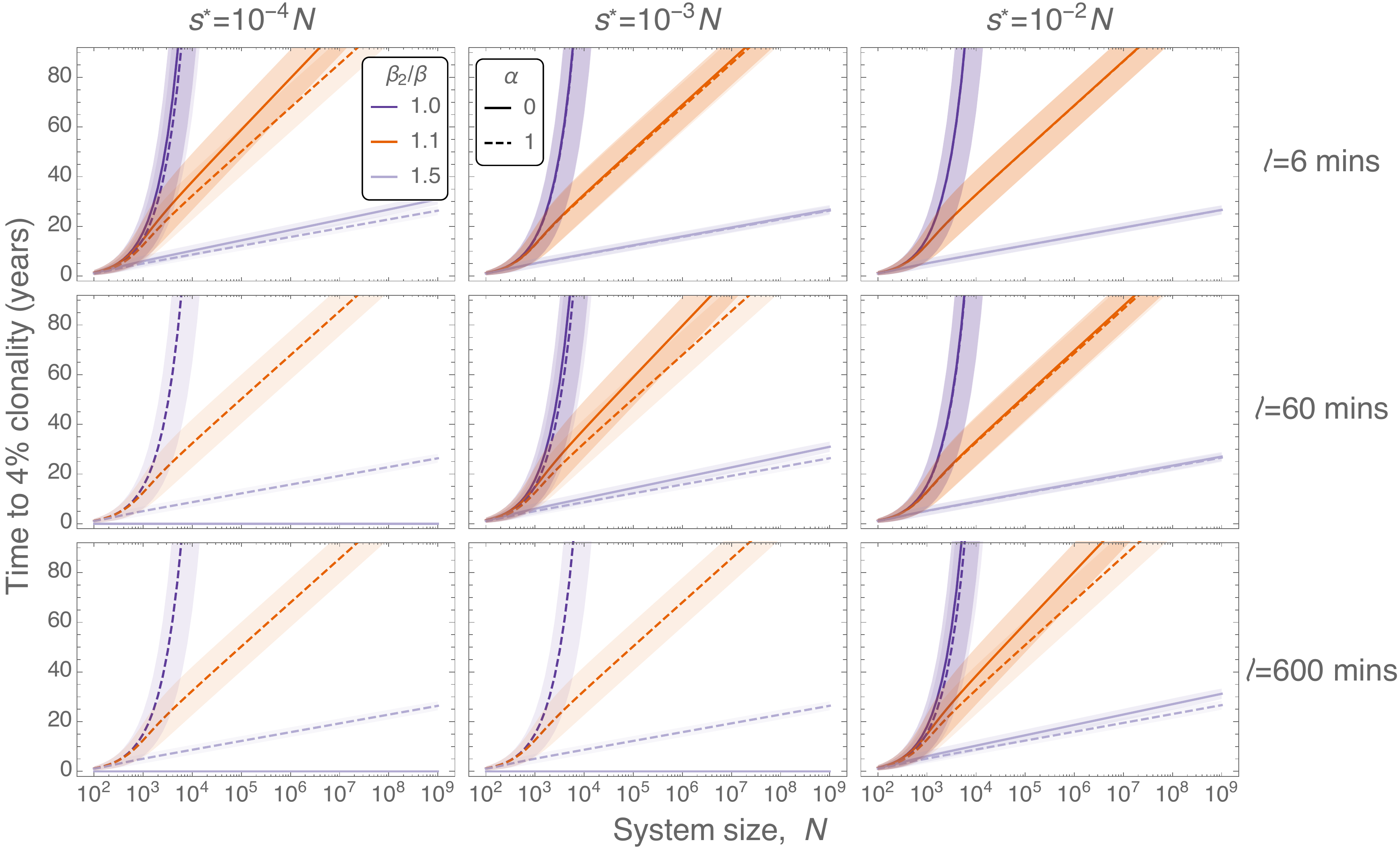}\end{figure*}}

\paragraph*{Fig~S1}
\label{fig:S1}
{\bf Clonality in man: more parameter combinations and death within niches.}
Time taken until a clone initiated from a single cell represents $4\%$ \cite{steensma:Blood:2015,sperling:NatRevCancer:2017} of the human HSC pool, as a function of the total number of niches in the system.
Colours represent the selective advantage of the invading clone.
Solid lines correspond to death only within the niches ($\alpha=0$), while dashed lines represent equal death rates in both compartments ($\alpha=1$; see SI for details).
Lines are generated using mathematical formulae in the SI.
Remaining parameters are $\beta=1/40$ week$^{-1}$ \cite{catlin:Blood:2011}, $n^*=0.99N$, and $\varrho=0$.
Some predictions are missing when $d \le 0$ and/or $a \le 0$; these parameter regimes are incompatible with our model.

\iftoggle{showSuppFigs}{\begin{figure*}[h]\centering\includegraphics[width=\textwidth]{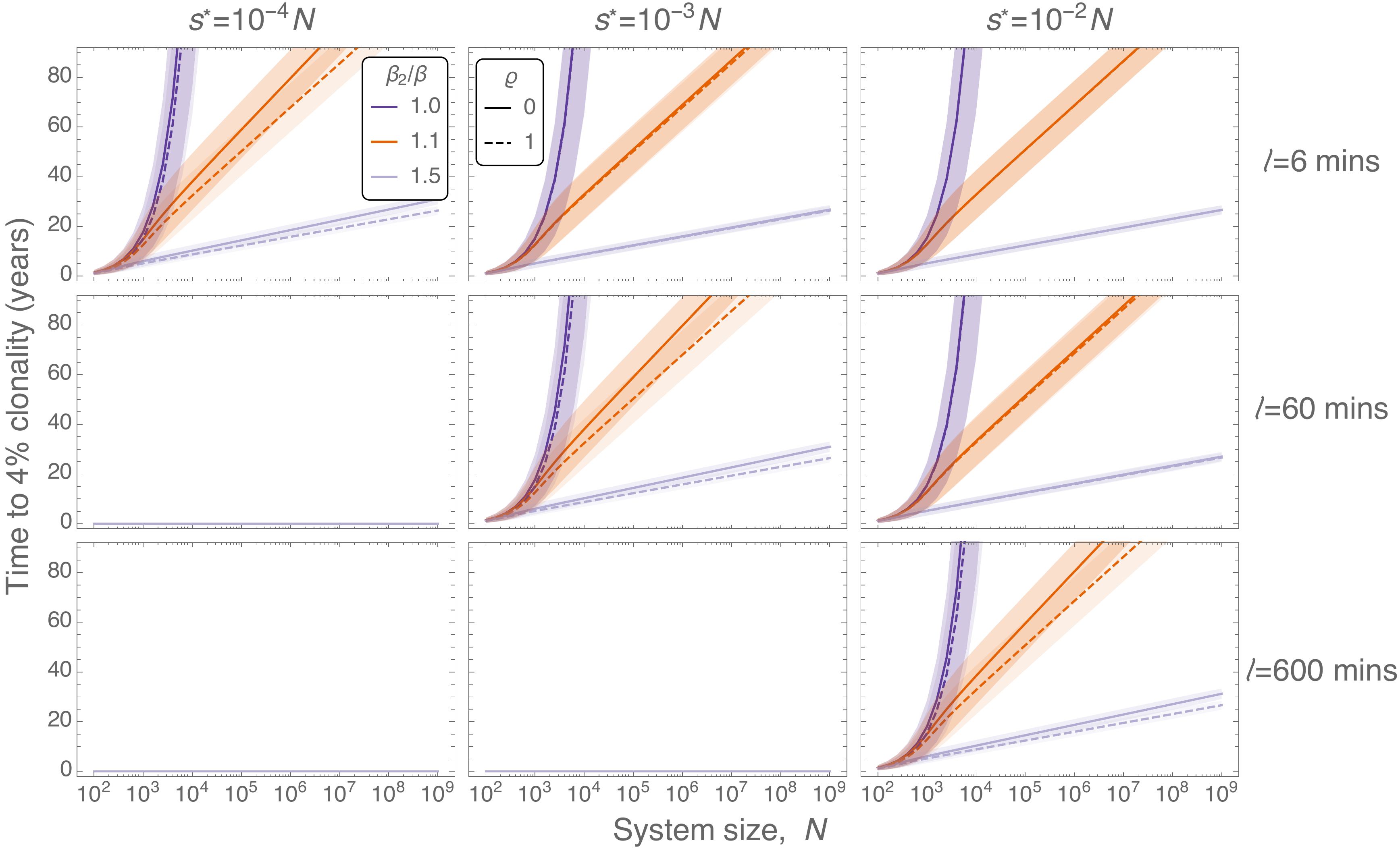}\end{figure*}}

\paragraph*{Fig~S2}
\label{fig:S2}
{\bf Clonality in man: more parameter combinations and reproduction into BM.}
Time taken until a clone initiated from a single cell represents $4\%$ \cite{steensma:Blood:2015,sperling:NatRevCancer:2017} of the human HSC pool, as a function of the total number of niches in the system.
Colours represent the selective advantage of the invading clone.
Solid lines correspond to the daughter cell entering the PB compartment after reproduction ($\varrho=0$), while dashed lines represent daughter cells remaining in the BM ($\varrho=1$; see SI for details).
Lines are generated using mathematical formulae in the SI.
Remaining parameters are $\beta=1/40$ week$^{-1}$ \cite{catlin:Blood:2011}, $n^*=0.99N$, and $\alpha=0$.
Some predictions are missing when $d \le 0$ and/or $a \le 0$; these parameter regimes are incompatible with our model.

\iftoggle{showSuppFigs}{\begin{figure*}[h]\centering\includegraphics[width=0.7\textwidth]{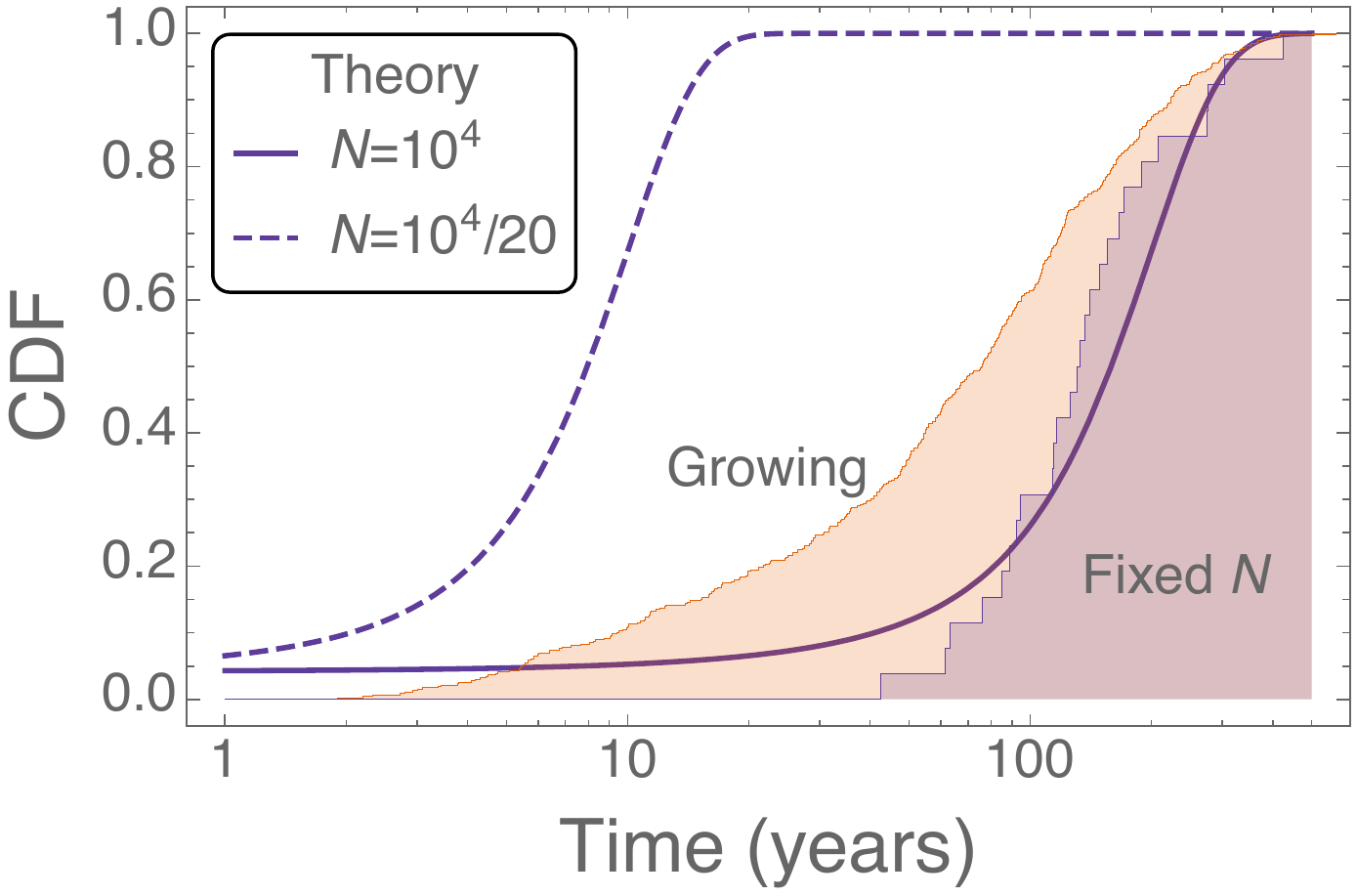}\end{figure*}}

\paragraph*{Fig~S3}
\label{fig:S3}
{\bf Predicted and simulated incidence curves of clonal hematopoiesis in man: constant size and under growth.}
The cumulative probability density function (CDF) of times to reach $4\%$ clonality \cite{steensma:Blood:2015,sperling:NatRevCancer:2017} when starting from a single neutral mutant in a normal host.
Shaded regions are incidence curves from simulations using either a constant niche count of $N=10^4$, or a logistically growing number of niches with $\dot{N} \approx r N(1-N/K)$, where $K=10^4$, $r=0.3$ per year, and $N(0)=K/20$.
These parameters represent a maturation period of $\sim 20$ years to reach $N \approx K$.
Lines are predicted incidence curves which assume normally-distributed times to clonality, using the mean and variance formulae as described in the SI, and constant population size as indicated in the legend (minimum and maximum number of niches).
Remaining parameters are $\beta=1/40$ week$^{-1}$ \cite{catlin:Blood:2011}, $n^*=0.99N$, $s^*=0.01N$, and $\ell=60$ minutes.
Finally, we only consider here the conditional incidence time, which have been normalised by the fixation probability.
This probability is 20 times larger for the neutral mutant in the growing model when compared to the fixed number of niches.

\section*{Acknowledgements}
The authors would like to thank Radek Skoda, Timm Schroeder, Ivan Martin, Larisa Kovtonyuk and Matthias Wilk for useful discussions.


\clearpage
\section*{Supporting Information (SI)}

\setcounter{section}{0}
\setcounter{equation}{0}
\setcounter{figure}{0}
\setcounter{table}{0}
\makeatletter
\renewcommand{\thesection}{SI.\Roman{section}}
\renewcommand{\theequation}{SI.\arabic{equation}}
\renewcommand{\thefigure}{SI.\arabic{figure}}
\renewcommand{\thetable}{SI.\arabic{table}}


\section{Model description}
\label{SI:sec:model}

The full model, as depicted in Fig 1 of the main manuscript, consists of four sub-populations and six processes for each cell type (wildtype and mutant/donor).
The number of host or wildtype cells located in the bone marrow (BM) is $n_1$, while $s_1$ is the number of cells of this type in the peripheral blood (PB).
Likewise, $n_2$ and $s_2$ are the number of mutant/donor cells in the BM and PB, respectively.
The BM has a maximum capacity of $N$ cells, representing the finite niche space in an organism.
The cell numbers are affected by birth, death, detachment from the BM, and attachment to the BM.
The effect of these events and the rate at which they happen are given by the following reactions:
\begin{linenomath}
\begin{subequations}
\label{SI:eq:reactions}%
\begin{align}
\mbox{Reproduction into PB:} \quad (n_i, s_i) &\xrightarrow{\makebox[7em]{$[1-\rho_i(n)] \beta_i n_i$}} (n_i, s_i+1), \\
\mbox{Reproduction into BM:} \quad (n_i, s_i) &\xrightarrow{\makebox[7em]{$\rho_i(n) \beta_i n_i$}} (n_i+1, s_i), \\
\mbox{Death in PB:} \quad (n_i, s_i) &\xrightarrow{\makebox[7em]{$\delta_i s_i$}} (n_i, s_i-1), \\
\mbox{Death in BM:} \quad (n_i, s_i) &\xrightarrow{\makebox[7em]{$\delta'_i n_i$}} (n_i-1, s_i), \\
\mbox{Detachment:} \quad (n_i, s_i) &\xrightarrow{\makebox[7em]{$d_i n_i$}} (n_i-1, s_i+1), \\
\mbox{Attachment:} \quad (n_i, s_i) &\xrightarrow{\makebox[7em]{$a_i s_i (N-n)/N$}} (n_i+1, s_i-1),
\end{align}
\end{subequations}
\end{linenomath}
where $n = \sum_i n_i$, and $(N-n)/N$ is the fraction of unoccupied niches.
The function $\rho_i(n)$ represents the probability for the new daughter cell following a reproduction event to attach directly to the BM, rather than entering the PB.
This function should satisfy $0 \le \rho_i(n) \le 1$, as well as $\rho_i(N)=0$.
For simplicity we choose a binary function such that 
\begin{equation}
\rho_i(n) = \left\{
\begin{matrix}
\varrho_i & \mbox{if } n < N, \\
0 & \mbox{otherwise.}
\end{matrix}\right.
\end{equation}
We express the death rate in the BM, $\delta'_i$, in terms of the original death rate $\delta_i$, such that $\delta'_i = \alpha \delta_i$.
With this parametrisation, setting $\alpha = 0$ prevents death from occurring within the BM, and $\alpha= 1$ makes HSC death independent of the environment.
As the BM is a more favourable environment for the HSCs, we expect $0 \le \alpha \le 1$.
As $\alpha$ is effectively a property of the environment, we assume it is identical for both host and mutant/donor cells.
We note here that there are no direct interaction terms between host cells and mutant/donor cells (no switching from type 1 to type 2 etc.).
We use the following parametrisation for the reaction parameters:
\begin{linenomath}
\begin{subequations}
\label{SI:eq:two-pop-parameters}%
\begin{align}
\beta_1	&= \beta,		&& \beta_2 	= (1+\varepsilon\gamma_\beta) \beta, \\
\delta_1	&= \delta,		&& \delta_2 	= (1+\varepsilon\gamma_\delta) \delta, \\
\delta'_1&= \alpha \delta_1 && \delta'_2 = \alpha \delta_2, \\
d_1		&= d,			&& d_2		= (1+\varepsilon\gamma_d) d, \\
a_1		&= a,			&& a_2 		= (1+\varepsilon\gamma_a) a, \\
\varrho_1 &= \varrho		&& \varrho_2 = (1+\varepsilon\gamma_\varrho) \varrho.
\end{align}
\end{subequations}
\end{linenomath}
Here $\varepsilon$ represents the strength of selection.
For analysis purposes discussed below, we assume $0 \le \varepsilon \ll 1$.
For $\varepsilon=0$, we have the so-called neutral model.
The parameters $\gamma_j$ ($j \in \{\beta,\delta,d,a,\varrho\}$) satisfy $\gamma_j = \mathcal{O}(1)$.
They allow the parameters to be varied independently.
In the main manuscript we restrict these parameters to $\gamma_\beta=1$ and $\gamma_\delta = \gamma_d = \gamma_a = \gamma_\varrho = 0$, along with $\alpha = \varrho = 0$, but we carry out the analysis in general here.

Considering a steady-state system in the absence of the mutant/donor cells ($n_2=s_2=0$), the deterministic dynamics of the system are described by ordinary differential equations (ODEs):
\begin{linenomath}
\begin{subequations}
\label{SI:eq:ODEs}%
\begin{align}
\frac{\dd n_1}{\dd t} &= (\varrho\beta - d -\alpha\delta) n_1 + a s_1 \frac{N-n_1}{N}, \label{SI:eq:ODEs-a} \\
\frac{\dd s_1}{\dd t} &= [d+(1-\varrho)\beta]n_1 - \left(\delta + a \frac{N-n_1}{N}\right)s_1 .
\end{align}
\end{subequations}
\end{linenomath}
From these ODEs we can obtain expressions for the equilibrium size of the BM and PB compartments, $n^*$ and $s^*$, respectively.
These are given by
\begin{linenomath}
\begin{subequations}
\label{SI:eq:equilibrium}%
\begin{align}
n^* &= N \left(1 - \frac{\delta(d+\alpha\delta-\varrho\beta)}{a(\beta-\alpha\delta)}\right) , \\
s^* &= \frac{(\beta-\alpha\delta)}{\delta} n^*. \label{SI:eq:equilibrium-b}
\end{align}
\end{subequations}
\end{linenomath}
From \eqref{SI:eq:equilibrium-b}, the death rate $\delta$ can be expressed in terms of $\beta$, $s^*$, $n^*$, and the variable parameter $\alpha$.
Furthermore, a cell in the PB compartment (in equilibrium) can either die with rate $\delta$ or attach to an unoccupied niche with rate $a(N-n^*)/N$.
Hence the expected lifetime of a cell in the PB is $\ell = [\delta + a(N-n^*)/N]^{-1}$.
From this we can obtain and expression for the attachment rate $a$.
Finally, $d$ is found from \eqref{SI:eq:ODEs-a}.
Thus for the non-valued parameters $\delta$, $d$, and $a$, we have
\begin{linenomath}
\begin{subequations}
\label{SI:eq:paramsub}%
\begin{align}
\delta &= \frac{\beta n^*}{s^* + \alpha n^*}, \\
d &= \frac{s^*}{\ell n^*} - (1-\varrho)\beta, \\
a &= \left(\frac{1}{\ell} - \frac{\beta n^*}{s^*+ \alpha n^*}\right)\frac{N}{N-n^*}
\end{align}
\end{subequations}
\end{linenomath}
These expressions, and the possible range of values, are given in Table~2 of the main manuscript for $\alpha = \varrho = 0$.
A further example is provided in Table~\ref{SI:tab:params3} of this document for $\alpha \ne 0$.
By introducing death in the niche, we affect the balance of cells leaving and entering each compartment.
The total number of cells produced ($\beta n^*$) must be matched by the total number of cells that die.
As we have increased the number of cells that are susceptible to death, we must decrease the death rate $\delta$.
Hence, $\delta$ is a decreasing function of $\alpha$.
To replace the cells that die within the BM, we need to increase the flux of cells from the PB to the BM.
In other words, death in the BM must be compensated by an increased rate of migration between the PB and BM compartments.
Hence, we have that $a$ is an increasing function of $\alpha$.
The detachment rate $d$ is independent of $\alpha$, but is an increasing function of $\varrho$ to ensure enough cells migrate to the PB.

\begin{table*}
\caption{Deduced parameter value ranges (units are per day). Here we have fixed $\ell = 3$ minutes and $\varrho=0$.}
\label{SI:tab:params3}%
\begin{tabular}{c l r c c c c l c c c c}
\hline\noalign{\smallskip}
 			& 																						& 			& \multicolumn{9}{c}{Value (per day)} \\
Parameter	& Expression 																			& $s^*$: 	& \multicolumn{4}{c}{1 cell} 							& \quad\quad	& \multicolumn{4}{c}{100 cells} \\
 			& 																						& $\alpha$:	& 0		& $10^{-4}$	& $10^{-2}$	& 1						& 			& 0	& $10^{-4}$							& $10^{-2}$	& 1 \\
\noalign{\smallskip}\hline\noalign{\smallskip}
$\delta$ 	& $\frac{\beta n^*}{s^* + \alpha n^*}$ 													& 			& 250	& 130		& 2.5		& 0.026 					& 			& \multicolumn{2}{c}{----- 2.5 -----}	& 1.3 		& 0.025 \\
\noalign{\smallskip}
$d$ 			& $\frac{s^*}{\ell n^*} - \beta$ 														& 			& \multicolumn{4}{c}{------------ 0.022 ------------}		& 			& \multicolumn{4}{c}{------------ 4.8 ------------} \\
\noalign{\smallskip}
$a$ 			& $\left(\frac{1}{\ell} - \frac{\beta n^*}{s^* + \alpha n^*}\right)\frac{N}{N-n^*}$	& 			& 23,000	& 35,000		& \multicolumn{2}{c}{--- 48,000 ---}	& 			& \multicolumn{4}{c}{--------- 48,000 ---------} \\
\noalign{\smallskip}\hline
\end{tabular}
\end{table*}

For completeness, the deterministic dynamics of the two-species system are described by the ODEs:
\begin{linenomath}
\begin{subequations}
\label{SI:eq:ODEsTwo}%
\begin{align}
\frac{\dd n_1}{\dd t} &= (\varrho\beta - d - \alpha \delta) n_1 + a s_1 \frac{N-n}{N}, \\
\frac{\dd n_2}{\dd t} &= (\varrho_2\beta_2 - d_2 - \alpha \delta_2) n_2 + a_2 s_2 \frac{N-n}{N}, \\
\frac{\dd s_1}{\dd t} &= [d + (1-\varrho)\beta]n_1 - \left(\delta + a \frac{N-n}{N}\right)s_1, \\
\frac{\dd s_2}{\dd t} &= [d_2 + (1-\varrho_2)\beta_2]n_2 - \left(\delta_2 + a_2 \frac{N-n}{N}\right)s_2,
\end{align}
\end{subequations}
\end{linenomath}
with $n=n_1+n_2$.

\section{Initial dynamics and chimerism}
\label{SI:sec:initial}
In the scenario of donor cell transplantation, the PB compartment initially contains more cells than the equilibrium value in the neutral model ($s_1 + s_2 > s^*$).
This leads to a net flux of cells attaching to the BM, and hence $n_1 + n_2 > n^*$.
The continuing attachment--detachment dynamics allows the donor cells to replace the host cells in the BM.
Meanwhile, surplus cells in the PB are dying off and the population relaxes to its equilibrium size ($n_1+n_2=n^*$ and $s_1+s_2 = s^*$).
The host cells in the BM are effectively displaced by the new donor HSCs.
Once the equilibrium is reached the initial dynamics end.
We find that the effect of selection, $\varepsilon$, only acts on a long timescale, and has little influence on the outcome of initial dynamics.
Therefore we treat the donor cells as neutral until the long-term noise-driven dynamics (discussed below) take over.

For small doses of donor HSCs, and especially the case of $\mathcal{S} = 1$ when a \emph{de novo} mutant is generated, the number of additional cells in the BM ($n_1+n_2-n^*$) is small.
We can neglect this expansion of the BM pool and approximate the number of occupied niches as a constant, $n^*$.
By considering only first-order reactions of the donor HSCs (i.e. the injected cells either die or attach to the BM, there is no reproduction or detachment), we can predict the number of donor HSCs that attach to the BM.
We find
\begin{equation}
n_2
= \frac{a(N-n^*)/N}{\delta + a(N-n^*)/N}\mathcal{S}
= \left(1-\frac{\ell \beta n^*}{s^*+\alpha n^*}\right)\mathcal{S}.
\label{SI:eq:chimerismLow}
\end{equation}
Hence for small doses the chimerism achieved is directly proportional to the dose size.
The simplified process that led to \eqref{SI:eq:chimerismLow} (only first-order dynamics) predicts $s_2 = 0$ after the initial dynamics, which is incorrect.
However, we know the relation between $s_i$ and $n_i$ in equilibrium [\eqref{SI:eq:equilibrium-b}], which tells us that $s_2 = (\beta - \alpha \delta) n_2 / \delta$.

If the dose of donor HSCs is large enough, all niches become occupied and the BM compartment is saturated.
In this case a niche that has been vacated (either by detachment or death) is immediately filled, either by a cell from the PB or from a reproduction event.
For example, if an $n_1$ cell detaches from the BM or dies within the BM, then it can be immediately replaced by an $s_2$ cell from the PB, or by an $n_2$ cell following reproduction with attachment.
Here we consider these combined detachment--attachment dynamics.
As the vacant niche is immediately occupied, we approximate the rate of the coupled detachment--attachment reaction as a single exponential step determined by the rate at which the niche becomes available (either $d$ or $\alpha\delta$).

For $s_1$ -- the number of wildtype cells in the PB -- we have an increase due to reproduction with rate $\beta n_1$ (niches are saturated so the new daughter cell enters the PB with certainty), and a decrease due to death at rate $\delta s_1$.
Furthermore, $s_1$ will increase if an $s_2$ attaches to the BM following the detachment of an $n_1$ cell; this process occurs at rate $d n_1 s_2/s$, where $s = s_1 + s_2$ and $s_2/s$ is the probability that an $s_2$ cell attaches rather than $s_1$.
Also, $s_1$ can decrease if it attaches to a niche vacated by $n_2$ cell.
Finally, an $s_1$ cell can attach to a niche that opens due to the death of the occupant, which happens with rate $\alpha \delta N s_1/s$ (here we assume $n_1+n_2=N$).
Similar dynamics follow for $s_2$.
Hence, for the PB cells we have the following simplified equations
\begin{linenomath}
\begin{subequations}
\label{SI:eq:ode:linear:s}%
\begin{align}
\frac{\dd s_1}{\dd t} &= \beta n_1 - \delta s_1 + d \left(n_1 \frac{s_2}{s} - n_2 \frac{s_1}{s}\right) - \alpha \delta N \frac{s_1}{s(t)}, \\
\frac{\dd s_2}{\dd t} &= \beta n_2 - \delta s_2 + d \left(n_2 \frac{s_1}{s} - n_1 \frac{s_2}{s}\right) - \alpha \delta N \frac{s_2}{s(t)}.
\end{align}
\end{subequations}
\end{linenomath}
The size of the PB pool follows the linear equation and solution
\begin{linenomath}
\begin{align}
\frac{\dd s}{\dd t} &= (\beta - \alpha \delta) N - \delta s \nonumber\\
\Rightarrow \quad s(t) &= \left( s(0) - \frac{(\beta - \alpha \delta) N}{\delta}\right)e^{-\delta t} + \frac{(\beta - \alpha \delta) N}{\delta}.
\label{SI:eq:neutralS}
\end{align}
\end{linenomath}
Here we assume the $N-n^*$ unoccupied niches are immediately filled after injection, such that $s(0) = s^* + \mathcal{S} - (N-n^*)$ and $n(t \ge 0) = N$.

For the cells in the BM, we construct the coupled dynamics in a similar way; a niche is cleared at rate $(d+\alpha\delta)N$, and the cell is replaced immediately by one from the PB, or by reproduction within the niche.
Wildtype BM cells can increase following the detachment/death of an $n_2$ cell with rate $(d+\alpha\delta)n_2[ (s_1/s) + (n_1/N) ]$.
Here $s_1/s$ represents the attachment of an $s_1$ cell (as opposed to $s_2$), and $n_1/N$ the reproduction within the niche of an $n_1$ cell.
The decrease of $n_1$ follows analogously, and occurs with rate $(d+\alpha\delta)n_1[ (s_2/s) + (n_2/N) ]$.
Hence, we can now write down the approximate ODEs for this scenario:
\begin{linenomath}
\begin{subequations}
\label{SI:eq:ode:linear:n}%
\begin{align}
\frac{\dd n_1}{\dd t} &= (d+\alpha\delta) \left(n_2 \frac{s_1}{s} - n_1 \frac{s_2}{s}\right), \\
\frac{\dd n_2}{\dd t} &= (d+\alpha\delta) \left(n_1 \frac{s_2}{s} - n_2 \frac{s_1}{s}\right),
\end{align}
\end{subequations}
\end{linenomath}
where the terms due to reproduction have cancelled out.
Although not obvious from above, \eqref{SI:eq:ode:linear:s} and \eqref{SI:eq:ode:linear:n} are actually linear in the $n_i$ and $s_i$.
Writing $n_1 = N-n_2$ and $s_1 = s - s_2$ we have
\begin{linenomath}
\begin{subequations}
\label{SI:eq:chimerismHigh}%
\begin{align}
\frac{\dd n_2}{\dd t} &= - (d+\alpha\delta)n_2 + \frac{(d+\alpha\delta)N}{s} s_2 , \\
\frac{\dd s_2}{\dd t} &= (\beta+d)n_2 - \left(\delta + \frac{(d+\alpha\delta)N}{s}\right)s_2 ,
\end{align}
\end{subequations}
\end{linenomath}
along with $n_2(0) = N-n^*$ and $s_2(0) = \mathcal{S} - (N-n^*)$.
Using these equations we can predict the value of $n_2$ once the PB has returned to its equilibrium size.

\section{Clonal dominance}
\label{SI:sec:dominance}

\subsection{General approach and the neutral scenario}

Once the mutant/donor cells have established themselves within the BM compartment, we want to know if and how quickly this clone expands.
To this end we use the projection method of Constable \emph{et al.} \cite{constable:PRE:2014,constable:PRL:2015}.
This analysis is more intuitive when the mutant/donor cells have no selective advantage, so we first discuss the neutral scenario.
In this case once the BM and PB compartments return to their equilibrium size ($n_1+n_2=n^*$ and $s_1+s_2=s^*$) there is no deterministic dynamics; as the mutant/donor cells are neutral when compared to the host, we have effectively returned to a healthy and stable host.
However, the stochastic dynamics of the individual-based model continue.
Cells are continually migrating between the BM and PB compartments, and reproduction and death events go on.
If cell numbers increase [decrease] then the flux of cells leaving the system increases [decreases] until the equilibrium is restored.
Therefore the deterministic dynamics constrains cell numbers to $n_1+n_2=n^*$ and $s_1+s_2=s^*$.
Cell number fluctuations, however, change the balance between host and mutant/donor cells over time, and we observe diffusion along the equilibrium line.
Eventually, this diffusion leads to the extinction of either the host or the mutant/donor population of HSCs.

We first move from the master equation -- the exact probabilistic description of the stochastic dynamics -- to a set of stochastic differential equations (SDEs) \cite{gardiner:book:2009}.
To this end we introduce the variables $\vec{x}=(s_1,s_2,n_1,n_2)^{\rm T}/N$ and expand the master equation in powers of $1/N$, using the fact that $N$ is a large parameter.
The evolution of $\vec{x}$ is determined by the set of SDEs
\begin{equation}
\frac{\dd x_i}{\dd t} = A_i(\vec{x}) + \frac{1}{\sqrt{N}} \eta_i(t),
\label{SI:eq:SDEsNeutral}
\end{equation}
where the $A_i$ are the drift terms representing the deterministic dynamics, and the $\eta_i$ are Gaussian noise terms which describe the diffusion.
The $\eta_i$ have zero expectation value and correlator
\begin{equation}
\langle \eta_i(t) \eta_j(t') \rangle = \delta(t-t')B_{ij}(\vec{x}).
\end{equation}
Here $\langle \cdot \rangle$ represents the expectation value over many realisations of the noise.
The exact forms of the drift vector $\vec{A}$ and diffusion matrix $\vec{B}$ (valid for $n<N$) are
\small
\begin{linenomath}
\begin{subequations}
\begin{align}
\vec{A}(\vec{x}) &= \begin{pmatrix}
[d+(1-\varrho)\beta] x_3 - (\delta + a y) x_1 \\
[d+(1-\varrho)\beta] x_4 - (\delta + a y) x_2 \\
(\varrho\beta-d-\alpha\delta)x_3 + a y x_1 \\
(\varrho\beta-d-\alpha\delta)x_4 + a y x_2
\end{pmatrix},\\
\vec{B}(\vec{x}) &= \begin{pmatrix}
[d+(1-\varrho)\beta] x_3 + (\delta + a y) x_1 & 0 & - d x_3 - a y x_1 & 0 \\
0 & [d+(1-\varrho)\beta] x_4 + (\delta + a y) x_2 & 0 & - d x_4 - a y x_2 \\
- d x_3 - a y x_1 & 0 & (\varrho\beta+d+\alpha\delta) x_3 + a y x_1 & 0 \\
0 & - d x_4 - a y x_2 & 0 & (\varrho\beta+d+\alpha\delta) x_4 + a y x_2 \\
\end{pmatrix}, \label{SI:eq:noiseMatrix}
\end{align}
\end{subequations}
\end{linenomath}
\normalsize
where we have used the shorthand $y=1-x_3-x_4$, which is the fraction of unoccupied niches.
\eqref{SI:eq:SDEsNeutral} is an approximate description of the full stochastic dynamics.
If we neglect the noise term, we recover the ODEs \eqref{SI:eq:ODEsTwo}.

The set of points at which the drift vector $\vec{A}$ is zero is known in dynamical systems theory as the slow manifold.
Our slow manifold, $\vec{x}^*$, satisfies the conditions $x_1^*+x_2^* = s^*/N$ and $x_3^*+x_4^* = n^*/N$, as well as $x_1^* = [(\beta-\alpha\delta)/\delta]x_3^*$ and $x_2^* = [(\beta-\alpha\delta)/\delta]x_4^*$.
The first two conditions describe the equilibrium size of the PB and BM compartments, and the latter conditions describe the balance between reproduction and cell death.
These four conditions can be satisfied parametrically by
\begin{equation}
\vec{x}^*(z) = \left( \frac{\beta-\alpha\delta}{\delta}(\xi-z), ~\frac{\beta-\alpha\delta}{\delta}z, ~\xi-z, ~z \right)^{\rm T},
\end{equation}
where $\xi=n^*/N$ and $z \in [0,\xi]$.
Hence, our slow manifold is a line through the 4-dimensional state-space.
In other words, if we were to measure the number of donor cells in the BM then we could infer the number of host cells from the system-size constraint.
These numbers, along with knowledge of the reproduction and death rates, can be used to infer cell numbers in the PB.
Therefore we only need to keep track of one variable, $z$, to describe our system.

As there is no deterministic drift along our slow manifold (in the neutral scenario), the time-evolution of the parametric coordinate $z$ satisfies
\begin{equation}
\frac{\dd z}{\dd t} = \frac{1}{\sqrt{N}} \eta(t),
\end{equation}
where $\eta(t)$ is Gaussian noise with zero expectation value and correlator
\begin{equation}
\langle \eta(t) \eta(t') \rangle = \delta(t-t') \widetilde{B}_{11}(z).
\end{equation}
The expansion (or contraction) of the mutant/donor clone is completely specified by this noise correlator $\widetilde{B}_{11}$, but it is not yet determined.
To find it we must project the approximate dynamics [\eqref{SI:eq:SDEsNeutral}] onto our slow manifold $\vec{x}^*$ \cite{constable:PRE:2014,constable:PRL:2015}.
In this way we capture the effects of the cell number fluctuations described above.
To achieve this we note that the Jacobian matrix of the drift vector along the slow manifold, $\vec{A}(\vec{x^*})$, has a zero eigenvalue.
This corresponds to the direction in which there is no deterministic motion, and hence the associated eigenvector is directed along the slow manifold.
Thus we use the eigenvectors of $\vec{A}(\vec{x^*})$ as a basis, onto which we decompose the SDEs~\eqref{SI:eq:SDEsNeutral}.
Selecting only the component along the slow manifold, we find the correlator
\begin{equation}
\widetilde{B}_{11}(z) = 2\mathcal{B} \,z(\xi-z),
\label{SI:eq:correlator}
\end{equation}
where the constant $\mathcal{B}$ is given by
\begin{linenomath}
\begin{align}
\mathcal{B}
&= \frac{\beta [d+(1-\varrho)\beta] [d+\alpha\delta(1-\varrho)] \delta^2 }{\xi \{\beta\delta + \beta d + d\delta + \alpha\delta[\beta-(d+\alpha\delta)] - \varrho\beta[\beta+\delta-\alpha\delta] \}^2} \nonumber\\
&=\frac{\beta n^* N(s^*+\alpha n^*)[s^*+\alpha n^*-n^*\ell(1-\varrho)\beta]}{[(n^*+s^*)(s^*+\alpha n^*)-n^*s^*\ell\beta]^2}.
\label{SI:eq:diffusionConstant}
\end{align}
\end{linenomath}
The first line of this equation expresses the diffusion constant in terms of the reaction parameters, while the second line uses \eqref{SI:eq:paramsub} to express $\mathcal{B}$ in terms of the experimentally-observed parameters, as well as $\alpha$ and $\varrho$.

By rescaling time and the coordinate $z$ in \eqref{SI:eq:correlator}, it can be shown that the dynamics along the slow manifold are equivalent to those of the Moran model \cite{constable:PRE:2014,constable:PRL:2015}.

We can use the standard results of Brownian motion to determine the probability that the mutant/donor clone expands to a given fraction $\sigma$, and the mean time for this to happen \cite{gardiner:book:2009}.
For example, $\sigma=0.5$ corresponds to a clone that represents $50\%$ of all HSCs.
We assume the dynamics starts at a point $\vec{x}^*(z_0)$ along the slow manifold.
Here $z_0=n_2/N$, where $n_2$ is the number of mutant/donor cells that make up the BM compartment at the end of the initial dynamics as described in the previous section.
In particular, for the case of disease spread ($\mathcal{S}=1$), $z_0$ can be found explicitly from \eqref{SI:eq:chimerismLow}.
The probability that the mutant/donor HSCs reach a fraction $\sigma \le 1$ is given by
\begin{equation}
\phi(z_0, \sigma) = \frac{z_0}{\sigma \xi}.
\label{SI:eq:neutralDominanceProb}
\end{equation}
The mean time for this expansion (i.e. the mean conditional time) is given by
\begin{equation}
T_\xi(z_0, \sigma) = \frac{N}{\mathcal{B}} \left[ \frac{\xi-z_0}{z_0} \log\left(\frac{\xi}{\xi-z_0}\right) + \frac{1-\sigma}{\sigma} \log(1-\sigma)\right].
\label{SI:eq:neutralDominanceTime}
\end{equation}
For $\sigma=1$, we recover the fixation probability and mean conditional fixation time of the mutant/donor cells.

\subsection{With selection}

We now repeat this analysis for the non-neutral case, i.e. the mutant/donor cells have a selective (dis)advantage.
For $\varepsilon>0$, the drift vector in \eqref{SI:eq:SDEsNeutral} becomes
\small
\begin{equation}
\vec{A}^{(\varepsilon)}(\vec{x})
=
\begin{pmatrix}
[d + (1-\varrho)\beta] x_3 - (\delta + a y) x_1 \\
[d + (1-\varrho)\beta] x_4 - (\delta + a y) x_2 \\
(\varrho\beta-d-\alpha\delta) x_3 + a y x_1 \\
(\varrho\beta-d-\alpha\delta) x_4 + a y x_2 
\end{pmatrix} 
+ \varepsilon
\begin{pmatrix}
0 \\
[d\gamma_d + (1-\varrho)\beta\gamma_\beta-\varrho\beta\gamma_\varrho] x_4 - (\delta\gamma_\delta + a\gamma_a y) x_2 \\
0 \\
[\varrho\beta(\gamma_\varrho+\gamma_\beta) - d\gamma_d - \alpha\delta\gamma_\delta]x_4 + a\gamma_a y x_2
\end{pmatrix}
+\mathcal{O}(\varepsilon^2).
\end{equation}
\normalsize
We assume there is no change in the noise correlator as terms $\mathcal{O}(\varepsilon/N)$ are negligible; hence we have $\vec{B}^{(\varepsilon)}(\vec{x}) \approx \vec{B}^{(0)}\vec{(x)} = \vec{B}(\vec{x})$, as given in \eqref{SI:eq:noiseMatrix}.

\begin{figure}
\centering
\includegraphics[width=0.7\textwidth]{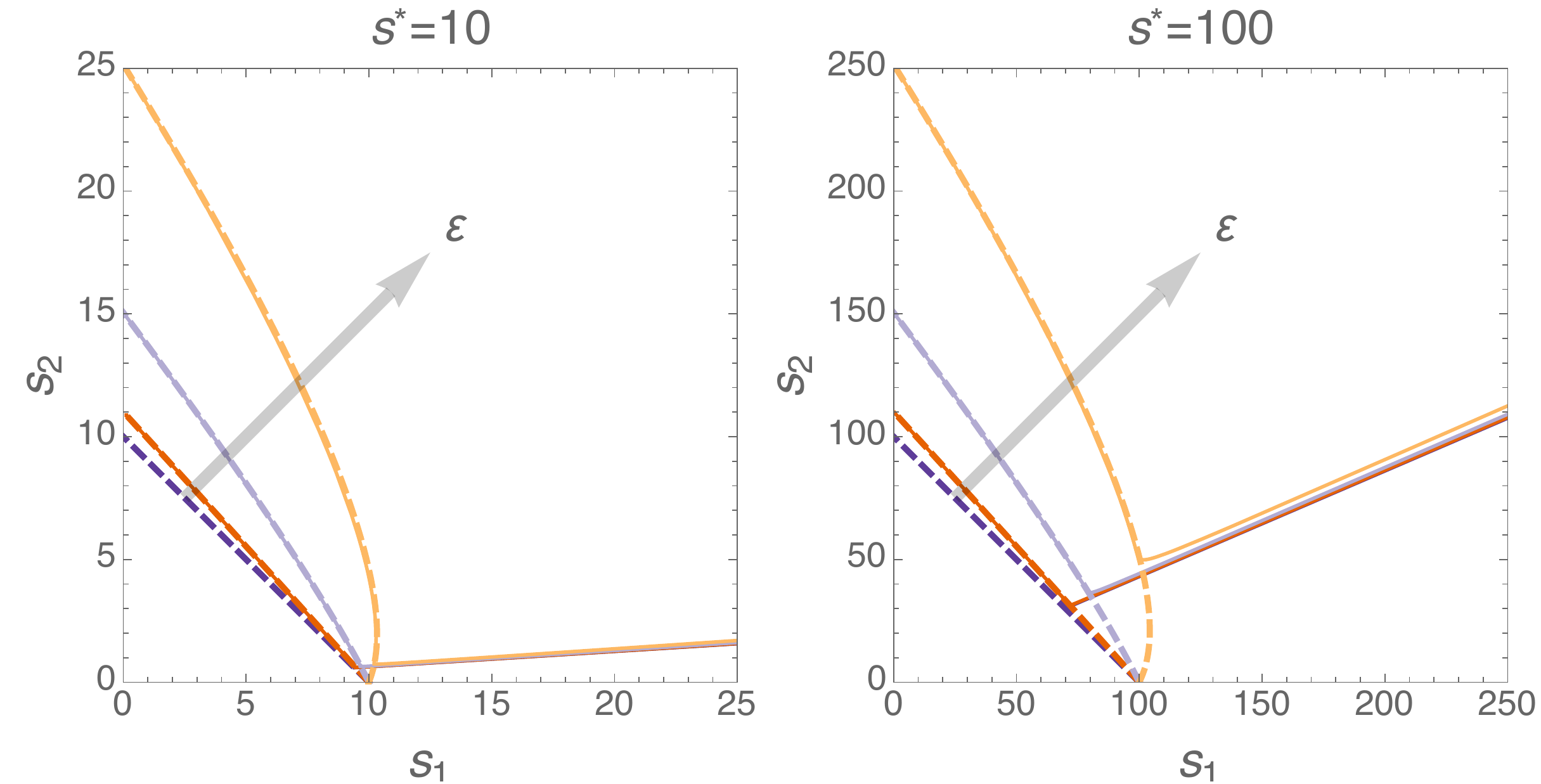}
\caption{
Time course of ODEs~\eqref{SI:eq:ODEsTwo} (solid lines) with an initial dose of $\mathcal{S} = 5,000$ donor cells into a steady-state host.
Selection strengths in this figure are $\varepsilon \in \{0,0.1,0.5,1.5\}$ (increasing in direction of arrow).
Dashed lines are the approximate slow subspaces, $\widetilde{\vec{x}}$, recovered from the projection method.
For $\varepsilon = 0$ the ODE time course stops once it reaches the slow manifold.
Increasing $\varepsilon$ moves the slow subspace away from the $\varepsilon=0$ slow manifold.
Here we have selection acting only on the reproduction rate ($\gamma_\beta=1$, $\gamma_a = \gamma_\delta = \gamma_d =0$), and we have $\ell = 3$ minutes.
In the left panel $s^* = 10$ and in the right panel $s^* = 100$.
We here set $\alpha=\varrho=0$.
Remaining parameters are as in Table~1 of the main manuscript.
}
\label{SI:fig:slowSubspace}
\end{figure}

As there is always some deterministic drift due to the effect of selection, by definition there is no slow manifold.
However, the number of cells leaving the system will still be balanced by production as described above.
This balance point describes a subspace around which the cell numbers fluctuate.
With $\varepsilon > 0$ there will be a tendency for the advantageous cells to replace their counterparts.
This induces a slow drift along the subspace.
By integrating the ODEs~\eqref{SI:eq:ODEsTwo} for a long time we can visualise this subspace.
Examples are shown in \figref{SI:fig:slowSubspace} for different selection strengths.
In the absence of selection ($\varepsilon = 0$), the trajectory from the ODEs~\eqref{SI:eq:ODEsTwo} stops once the slow manifold has been reached.
This is the reason why the dashed line is not accompanied by a solid trajectory for $\varepsilon = 0$.
For $\varepsilon > 0$, the advantageous mutant/donor cells are able to maintain a higher equilibrium population size than the host.
Hence the slow subspaces always lie above the neutral slow manifold.

We can use the projection method to calculate the approximate form of the slow subspace, $\widetilde{\vec{x}}(z)$ \cite{constable:PRE:2014,constable:PRL:2015}.
This takes the form $\widetilde{\vec{x}}(z) = \vec{x}^*(z) + \varepsilon f(z)$, and again we only need one variable to describe our system.
These approximations are shown as dashed lines in \figref{SI:fig:slowSubspace}, and they remain highly accurate (compared to numerical integration of the ODEs~\eqref{SI:eq:ODEsTwo}) even for large values of $\varepsilon$.

Using the same eigenbasis from the neutral model we can project the dynamics onto the slow subspace \cite{constable:PRE:2014,constable:PRL:2015}.
The SDE describing the motion along the slow subspace is
\begin{linenomath}
\begin{align}
\frac{\dd z}{\dd t} &= \widetilde{A}_1(z) + \frac{1}{\sqrt{N}} \eta(t), \nonumber\\
\langle \eta(t) \rangle &= 0, \nonumber\\
\langle \eta(t)\eta(t') \rangle &= \delta(t-t')\widetilde{B}_{11}(z).
\end{align}
\end{linenomath}
The drift along this subspace, $\widetilde{A}_1(z)$, is of the form
\begin{equation}
\widetilde{A}_1(z) = \varepsilon \mathcal{A} \, z(\xi-z),
\end{equation}
where the constant $\mathcal{A}$ is given by
\small
\begin{linenomath}
\begin{align}
\mathcal{A} 
&= \frac{\delta}{\xi}\frac{\gamma_a(\beta-\alpha\delta)(d+\alpha\delta-\varrho\beta) - \gamma_d d(\beta-\alpha\delta) + \gamma_\beta \beta[d+\alpha\delta(1-\varrho)] - \gamma_\delta[\alpha\delta(\beta-\alpha\delta)-\beta(\varrho\beta-d-\alpha\delta)]+\gamma_\varrho\varrho\beta(\beta-\alpha\delta)]}{\beta\delta+\beta d+d\delta+\alpha\delta(\beta-d-\alpha\delta)-\varrho\beta[\beta+(1-\alpha)\delta]}.
\end{align}
\end{linenomath}
\normalsize
By setting $\alpha=\varrho=0$, we recover
\begin{equation}
\mathcal{A} = \frac{d \beta \delta}{\xi (d\beta + d\delta + \beta\delta)}( \gamma_\beta + \gamma_a - \gamma_\delta - \gamma_d ).
\end{equation}
This expression gives an important result; it doesn't matter in which sense the cells are advantageous.
A cell with an increased reproduction rate ($\gamma_\beta = 1$; $\gamma_a = \gamma_\delta = \gamma_d = 0$) has the same advantage as a cell which attaches to the BM more quickly ($\gamma_a = 1$; $\gamma_\beta = \gamma_\delta = \gamma_d = 0$).
All that matters is the cumulative advantage, $\gamma_\beta + \gamma_a - \gamma_\delta - \gamma_d$.
For this reason we only consider a reproductive advantage in the presented results in the main manuscript.

\begin{figure}
\centering
\includegraphics[width=0.4\textwidth]{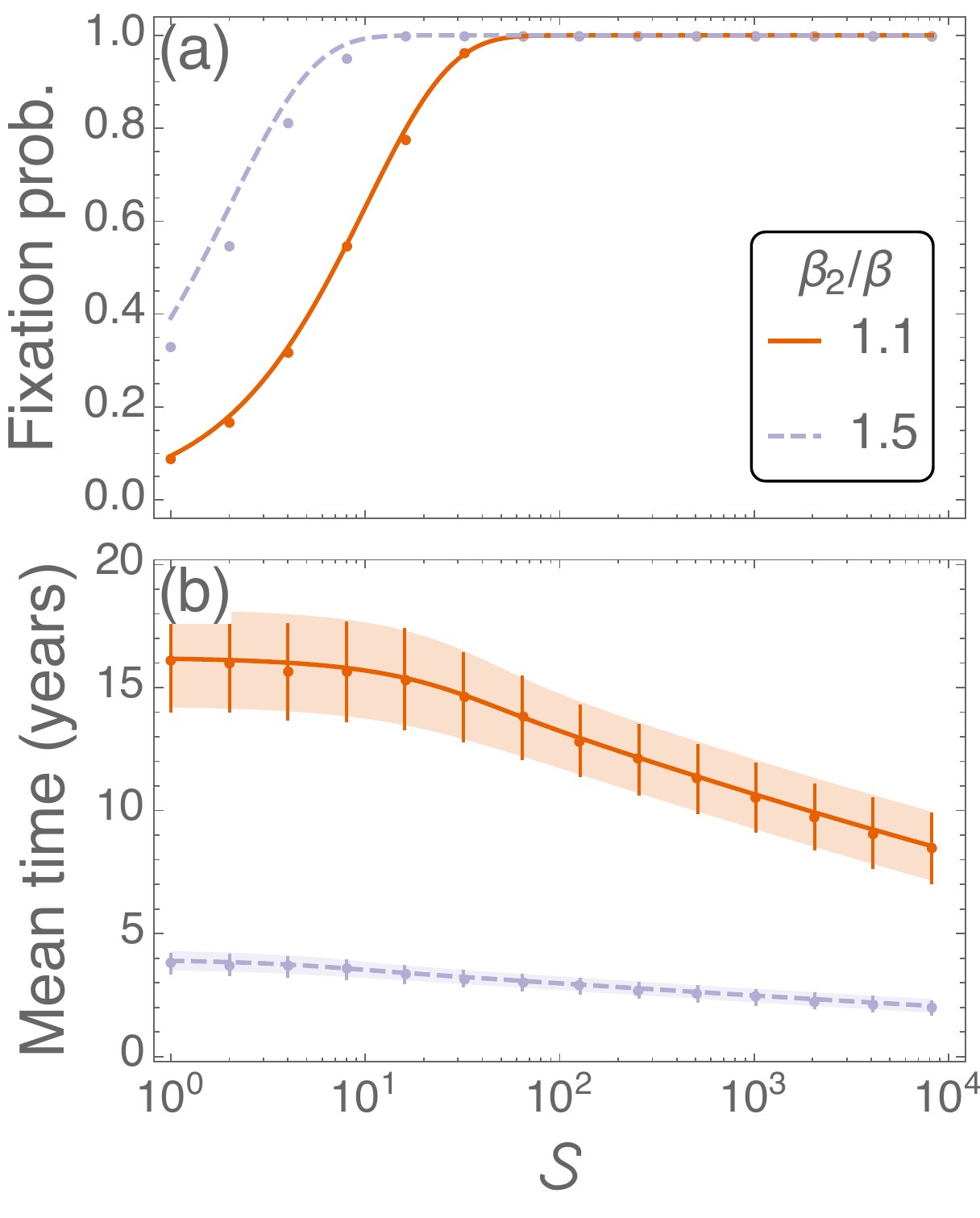}
\caption{
Fixation probability (a) and time (b) of donor cells in a non-preconditioned host.
On the horizontal axis we plot the initial dose of donor HSCs, $\mathcal{S} \in \{1,2,4,8,\dots,8192\}$.
Symbols are results from $10^3$ simulations of the stochastic model (with associated standard deviations). 
Lines are predictions from \eqref{SI:eq:selectiveDominanceProb} and \eqref{SI:eq:selectiveDominanceTime}.
The shaded region in (b) is the standard deviation calculated from the second moment, \eqref{SI:eq:secondMoment}.
Here we have $s^* = 100$ cells, and $\ell = 3$ minutes, along with $\alpha=\varrho=0$.
Remaining parameters are as in Table~1 of the main manuscript.
}
\label{SI:fig:transplantProbTime}
\end{figure}

Again using the standard results of Brownian motion \cite{gardiner:book:2009}, we find the probability for the donor cells to represent a fraction $\sigma$ of the population to be
\begin{equation}
\phi(z_0, \sigma) = \frac{1 - e^{-\Lambda z_0}}{1 - e^{-\Lambda \sigma \xi}}, \qquad \mbox{with}\quad
\Lambda = \frac{\varepsilon N \mathcal{A}}{\mathcal{B}}.
\label{SI:eq:selectiveDominanceProb}
\end{equation}
This solution is of the same form as that obtained for a Moran model \cite{constable:PRE:2014,constable:PRL:2015}.
Although a closed-form solution is possible for the mean conditional time to reach size $\sigma$, it is too long to display here.
Instead we use an algebraic software package to solve the second-order differential equation
\begin{equation}
T_\xi(z_0, \sigma) = \frac{\theta(z_0, \sigma)}{\phi(z_0, \sigma)},
\qquad
\frac{\partial^2 \theta(z_0, \sigma)}{\partial z_0^2} + \Lambda \frac{\partial \theta(z_0, \sigma)}{\partial z_0} = -\frac{N}{\mathcal{B}} \frac{\phi(z_0, \sigma)}{z_0(\xi-z_0)}, \qquad \theta(0) = \theta(\sigma \xi) = 0.
\label{SI:eq:selectiveDominanceTime}
\end{equation}
Example results from \eqref{SI:eq:selectiveDominanceProb} and \eqref{SI:eq:selectiveDominanceTime} are shown in \figref{SI:fig:transplantProbTime}.

Furthermore, we can write down a set of equations for the second moment of the conditional time for the mutant/donor clone to reach size $\sigma$ \cite{goel:book:1974}.
The second moment, $\langle T_\xi^2(z_0,\sigma) \rangle$, is dependent on the first moment of the conditional time, and hence we must solve the coupled equations:
\begin{linenomath}
\begin{subequations}
\label{SI:eq:secondMoment}%
\begin{align}
\frac{\partial^2 \theta_1(z_0, \sigma)}{\partial z_0^2} + \Lambda \frac{\partial \theta_1(z_0, \sigma)}{\partial z_0} = -\frac{N}{\mathcal{B}} \frac{\phi(z_0, \sigma)}{z_0(\xi-z_0)},& \qquad \theta_1(0) = \theta_1(\sigma \xi) = 0, \\
\frac{\partial^2 \theta_2(z_0, \sigma)}{\partial z_0^2} + \Lambda \frac{\partial \theta_2(z_0, \sigma)}{\partial z_0} = -2\frac{N}{\mathcal{B}} \frac{\theta_1(z_0, \sigma)}{z_0(\xi-z_0)},& \qquad \theta_2(0) = \theta_2(\sigma \xi) = 0, \\
\langle T_\xi^2(z_0,\sigma) \rangle = \frac{\theta_2(z_0, \sigma)}{\phi(z_0, \sigma)},
\end{align}
\end{subequations}
\end{linenomath}
where the first equation is for the first moment [identical to \eqref{SI:eq:selectiveDominanceTime}], and the second equation is for the second moment.
The predicted deviation calculated from this second moment is shown in \figref{SI:fig:transplantProbTime}(b).

Full details of all these calculations are found in the accompanying Mathematica notebook file, which can be found at \url{https://github.com/ashcroftp/clonal-hematopoiesis-2017}.

\subsection{Further neutral-model calculations}
The expansion of neutral clones deserves some further attention, as this process could provide valuable insight into human hematopoiesis.
Firstly, we look at the case $\alpha=\varrho=0$ and ask how does the mean conditional time to $\sigma$-level clonality vary depending on the choice of model parameters.
Using \eqref{SI:eq:chimerismLow}, along with $\mathcal{S}=1$, we have an expression for our initial level of clonality:
\begin{equation}
z_0 = \frac{1}{N}\frac{s^* - n^* \beta \ell}{s^*}.
\end{equation}
From \eqref{SI:eq:diffusionConstant}, we also have the diffusion constant
\begin{equation}
\mathcal{B} = \frac{\beta N}{s^*} \frac{\frac{s^*}{n^*} - \beta \ell}{\left(1 + \frac{s^*}{n^*} - \beta \ell\right)^2}.
\end{equation}
Therefore, our mean time [\eqref{SI:eq:neutralDominanceTime}] satisfies
\begin{equation}
\langle T_\xi(z_0, \sigma) \rangle = \frac{s^*}{\beta} \frac{\left(1+ \frac{s^*}{n^*} - \beta \ell\right)^2}{\frac{s^*}{n^*} - \beta \ell} \left[ \left(\frac{n^* s^*}{s^* - n^* \beta \ell}-1\right)\log\left(\frac{1}{1-\frac{s^* - n^* \beta \ell}{n^* s^*}}\right) + \frac{1-\sigma}{\sigma} \log(1-\sigma)\right].
\end{equation}
Now assuming that terms $\mathcal{O}(N)$ are much larger than terms $\mathcal{O}(1)$, that $s^*/n^* \ll 1$ (blood compartment much smaller than bone marrow in equilibrium) and $\beta \ell \ll 1$ (migration dynamics are faster than reproduction), and that $\sigma \ll 1$, we can approximate our mean time as
\begin{equation}
\langle T_\xi(z_0, \sigma) \rangle \approx \frac{1}{\beta} \frac{\sigma}{2} \frac{n^* s^*}{s^* - n^* \beta \ell}.
\end{equation}
Note that we are also constrained to $s^* - n^* \beta \ell > 0$, which comes from the fact that our attachment and detachment parameters $a$ and $d$ must be positive.
If $s^* \gg n^* \beta \ell$, then we have $\langle T_\xi(z_0, \sigma)\rangle \approx \sigma n^* / (2\beta)$, which is independent of $\ell$ and $s^*$.
As expected, the reproduction rate $\beta$ is the dominant parameter determining the time to clonality.

When considering the case of $\alpha \ne 0$ and $\rho \ne 0$, we turn to a graphical representation to highlight the parameter dependence.
These results are shown in supplementary figures S1 and S2.
We find that allowing equal death in both compartments ($\alpha=1$) or daughter cells to enter the niche directly ($\varrho=1$) has little to no effect on the time to clonality.

Finally, we can obtain a closed-form solution of the second moment equation [\eqref{SI:eq:secondMoment}] in the absence of selection.
We find
\begin{equation}
\langle T_\xi^2(z_0,\sigma) \rangle = \frac{2 N}{\mathcal{B}}\left[ \langle T_\xi(z_0,\sigma) \rangle\left(\frac{1-\sigma}{\sigma}\log(1-\sigma)-1\right) + \frac{N}{\mathcal{B}}\left({\rm Li}_2(\sigma) - {\rm Li}_2\left(\frac{z_0}{\xi}\right) \right)\right],
\end{equation}
where ${\rm Li}_2(z)$ is the second-order polylogarithmic function.
\section{Engraftment into a preconditioned host}
\label{SI:sec:preconditioned}

Even if the BM compartment is empty, in the stochastic model there is a finite probability that all donor cells die before they engraft.
We write $\psi = {\rm Pr}(n+s=0, \,t\to\infty)$ for the extinction probability of a single cell.
Therefore, the probability that a single donor HSC will reconstitute the preconditioned host is $\varphi = 1-\psi = {\rm Pr}(n+s>0, \,t\to\infty)$.
For doses of size $\mathcal{S}$, the reconstitution probability is $\varphi = 1-\psi^{\mathcal{S}}$.
For a first-approximation of this probability, we assume that the donor cells can only either attach to the BM niches or die.
The probability that a single HSC in the PB compartment dies is $\psi = \delta/(\delta+a)$.
Thus the approximate reconstitution probability is
\begin{equation}
\varphi = 1 - \left(\frac{\delta}{\delta + a}\right)^\mathcal{S}.
\label{SI:eq:emptyReconst0}
\end{equation}
Here we have assumed that all niches are unoccupied, such that the attachment rate per cell is $a(N-0)/N=a$.

We can also consider multiple attachments and detachments, which could occur before a cell establishes a sustainable population or dies.
As above, the probability that a single HSC in the PB compartment dies is $\delta/(\delta+a)$.
Alternatively, the HSC can attach to the niche with probability $a/(\delta+a)$.
The probability that the cell then detaches from the niche is $d/(d+\beta+\alpha\delta)$, the probability of dying within the niche is $\alpha\delta/(d+\beta+\alpha\delta)$, and the probability of reproducing is $\beta/(d+\beta+\alpha\delta)$.
Again we have assumed that the number of occupied niches is negligible. 
Under these processes, the probability that a single donor HSC becomes extinct is given by
\begin{equation}
\psi = \frac{\delta}{\delta+a} + \frac{a}{\delta+a}f(\psi),
\label{SI:eq:emptyPsi}
\end{equation}
where $f(\psi)$ represents the processes that occur once the cell has entered the BM compartment.
It is given by
\begin{equation}
f(\psi) = \frac{\alpha\delta}{d+\beta+\alpha\delta} + \frac{d}{d+\beta+\alpha\delta} \psi + \frac{\beta}{d+\beta+\alpha\delta}\left[(1-\varrho)\psi f(\psi) + \varrho f^2(\psi)\right].
\label{SI:eq:emptyPsi2}
\end{equation}
Here the first term is the death of the cell within the BM compartment.
The second term represents detachment and the cell is back where it started, so this is multiplied by $\psi$.
The third term represents reproduction: either one offspring is ejected to the PB (hence $\psi$) and the other remains in the niche [$f(\psi)$], or both offspring remain in the BM [$f^2(\psi)$].
Solving \eqref{SI:eq:emptyPsi2} for $f(\psi)$, and then using this to solve \eqref{SI:eq:emptyPsi}, we find the extinction probability of a single cell.
For $\varrho=0$, this is simply
\begin{equation}
\psi = \frac{\delta}{\delta+a} \frac{d+\beta+\alpha(\delta+a)}{\beta},
\end{equation}
and hence the reconstitution probability given a dose of $\mathcal{S}$ donor HSCs is
\begin{equation}
\varphi = 1- \left(\frac{\delta}{\delta + a} \frac{d + \beta + \alpha(\delta + a)}{\beta}\right)^{\mathcal{S}}.
\label{SI:eq:new-Reconst}
\end{equation}

\end{document}